\documentclass[onecolumn]{aa}
\usepackage{graphicx}

\begin{document}

%
\title{Relativistic parsec--scale jets: \uppercase{ii}. Synchrotron emission}

\author
{V.I. Pariev\inst{1}\inst{2}
	\and
        Ya.N. Istomin\inst{1}
	\and
	A.R. Beresnyak\inst{1}}

	\offprints{Ya.N. Istomin}

	\institute{
P.~N.~Lebedev Physical Institute, 
Leninsky Prospect 53, Moscow 117924, Russia \\
email: istomin@td.lpi.ac.ru
    \and
Department of Physics and Astronomy, University of Rochester, 
Rochester, NY 14627, USA \\
email: vpariev@pas.rochester.edu
}

	\date{Received  ; accepted  }

\titlerunning{Relativistic parsec--scale jets: \uppercase{ii}}
\authorrunning{V.I. Pariev et al.}

\abstract{
We calculate the optically thin synchrotron emission
of fast electrons and positrons in a spiral stationary magnetic field
and a radial electric field of a rotating relativistic strongly magnetized 
force--free jet consisting of electron-positron pair plasma.
The magnetic field has a helical structure with a uniform 
axial component and a toroidal component that is maximal inside the 
jet and decreasing to zero 
towards the boundary of the jet. Doppler boosting and swing of the 
polarization angle of synchrotron emission
due to the relativistic motion 
of the emitting volume are calculated. The 
distribution of the plasma velocity
in the jet is consistent with the electromagnetic field structure.
Two spatial distributions of fast particles are considered: uniform, and 
concentrated in the vicinity of the Alfv\'en resonance surface.
The latter distribution corresponds to the regular acceleration
by an electromagnetic wave in the vicinity of its Alfv\'en 
resonance surface inside the jet. The polarization properties of the
radiation have been obtained and compared with the 
existing VLBI polarization measurements
of parsec--scale jets in BL~Lac sources and quasars.
Our results give a natural explanation of the observed
bimodality in the alignment between the electric field vector of 
the polarized radiation and the projection of the jet axis on the plane 
of the sky. We interpret the motion of bright knots as a phase velocity of 
standing spiral eigenmodes of electromagnetic perturbations
in a cylindrical jet.   
The degree of polarization and the velocity of the observed 
proper motion of bright knots
depend upon the angular rotational velocity of the jet. The observed 
polarizations and velocities of knots indicate that the magnetic field lines
are bent in the direction opposite to the direction of the jet rotation.

      }

\maketitle

\newcommand{\beg}{\begin{equation}}
\newcommand{\en}{\end{equation}}
\def\e{{\cal E}}
\def\fracb#1#2{\frac{\displaystyle #1}{\displaystyle #2}}
\def\pa{\partial}
\def\om{\omega}
\def\omb{{\omega_B}}
\def\der#1#2{\frac{{\pa #1}}{{\pa #2}}}
\def\der#1#2{\frac{{\pa #1}}{{\pa #2}}}
\def\g{\zeta}
\def\omr{\Omega^F r}
\def\chit{\tilde\chi}
\section{Introduction}\label{sec1}

The jets from Active Galactic Nuclei are not uniform. They consist 
of a number of bright knots with fainter emission in between.
It is well established that the nature of the radio emission from 
jets on kiloparsec and parsec scales is synchrotron radiation of 
relativistic nonthermal particles (Begelman et al. 1984).
There is mounting observational evidence 
that jets in many AGNs are electron-positron rather than electron-proton.
The spectral model of the jet
radiation in a broad frequency range (from X-rays to radio), is
described by Morrison \& Sadun (1992), who pointed out that the 3C273 jet must
be dominated by electron-positron pairs.
Reynolds et al. (1996) used early VLBI observations of the M87 jet
and self-absorbed core. They used the theory of synchrotron self-absorption
to put a constraint on the electron density in the core and concluded
that the core of M87 is likely to be e$^{+}$e$^{-}$ pair plasma.
A similar method was used by Hirotani et al. (1999) and Hirotani et al. (2000)
to reveal that the components in the jets of 3C279 and 3C345 are also
dominated by pair plasma.
Observations of circular
polarization of radio emission from 3C279, 3C273, 3C84 and
PKS0528+134 (Wardle et al., 1998) support the same conclusion.
However, there are also indications that jets in Optically
Violently Variable quasars cannot consist solely of e$^{+}$e$^{-}$
pairs because they would produce much larger soft X-ray
luminosities than are observed. On the other hand, models with
jets consisting solely of proton-electron plasma are excluded,
since they predict much weaker nonthermal X-radiation than
observed in Optically Violently Variable quasars (Sikora~\&
Madejski, 2000).
The {\it in situ} acceleration of relativistic particles seems unavoidable
because high synchrotron losses prevent relativistic particles from 
streaming out of the central engine to the distances of bright knots (Begelman
et al. 1984; R\"oser~\& Meisenheimer 1991).
There are several possible mechanisms of particle acceleration 
inside knots. The acceleration can be by shocks 
(Blandford~\& Eichler 1987), by 
turbulent electromagnetic fields (Manolakou, Anastasiadis~\& Vlahos 1999)   
and by magnetic reconnection 
(Romanova~\& Lovelace 1992; Blackman 1996). Another 
possible mechanism of particle acceleration is the acceleration by 
organized electromagnetic fields which are the eigenmodes of the jet as
a whole. These modes can be excited as a result of non-stationary processes 
in the central engine. The energy can be transmitted by such global waves along
the jet to large distances without noticeable attenuation. 

The acceleration
is most effective if resonances of the waves with the mean stream take place
(Beresnyak et al. 2002). 
The resonance with Alfv\'en waves occurs when $\omega^\prime=k_\|^\prime c$,
where $\omega^\prime$ and $k_{\parallel}^\prime$ are the frequency and
the projection of the wave vector on the magnetic field in the plasma rest frame.
When the specific energy density and pressure of the plasma 
are much less than the
energy density of the magnetic field, 
The Alfv\'en velocity is equal to the speed
of light in vacuum $c$. 
The resonance condition is fulfilled
on a specific magnetic surface and, in the case of a cylindrical jet, at
a definite distance from the axis. In the vicinity of that surface the magnetic
and electric fields of the wave reach large magnitudes. 
The resonances occur for nonaxisymmetric
modes of global waves and the position and width of the resonances depend 
on the structure of the ordered helical magnetic field. 
Both electrons and positrons are subject to
drifting out of the region of the strong electric field of the wave near the
resonance due to the non-uniformity of the electric field.
When moving to a
region of weak electric field, particles gain energy. This energy
is taken from the energy of the wave, which is an eigenmode in the
jet, and the eigenmode decays. 
Beresnyak at al. (2002) considered this mechanism of 
particle acceleration. Drift approximation was used to describe the
motion of a particle in the electric field of the wave.
Synchrotron losses and isotropization of the particle
distribution by plasma instabilities were taken into account. 
Plasma instabilities are
excited by accelerated particles if the distribution function is
anisotropic in momentum space, and act to conceal any anisotropy in
the particle distribution. Acceleration process and
synchrotron losses taken together yield a power law energy spectrum
of ultra-relativistic electrons and positrons with an index
varying between $2$ and $3$ depending upon the initial energy of the 
injected particles. This is consistent with the typical spectral 
indices of radio emission observed in parsec--scale jets.

The polarization of
synchrotron emission in knots is very sensitive to the geometry of the 
helical large scale magnetic field. Indeed, VLBI observations provide evidence
for the polarized emission in knots and in the space between knots (Gabuzda 1999a;
Hutchison et al. 2001).
This is indicative of the existence of a large scale magnetic field all over
the jet rather than concentrated in separate knots. In the general case 
a helical magnetic field has poloidal and toroidal components. The toroidal component
arises due to the axial electric current in the jet. If the current inside the jet is 
closed (no net jet electric current), that is the electric current flows in 
different directions at different radii in the jet, then the toroidal magnetic field
must be small at the boundary of the jet and only the axial component of the magnetic 
field will remain. The change of the direction of the magnetic field from toroidal
close to the jet axis to axial on the periphery was inferred from the 
VLBI polarization measurements in blazar 1055+018 (Attridge et al. 1999).
The concept of an intrinsic large-scale helical magnetic field is the simplest 
model of the jet and does not require additional processes of amplification or
generation of azimuthal and axial components of the magnetic field in shocks or 
shearing layers at the jet boundary.
   
The polarization of synchrotron emission of parsec scale jets is being studied
by few groups (Gabuzda 1999b, 2000; Lepp\"anen at al. 1995; 
Lister 2001). We use observational data to compare with our calculations of polarization 
in a model of a force--free strongly magnetized relativistic rotating jet. We
calculate the polarization of synchrotron radiation in the helical field
geometry (Sect.~\ref{sec2}). In Sect.~\ref{sec3} we expand our calculations 
to include the
Doppler boosting effect and swing of the polarization angle
due to relativistic motion and rotation of the plasma in the jet.
We also compare the observed apparent velocities of knots with the velocities 
given by our model of propagation of spiral perturbations in a cylindrical jet.
As it was shown by Istomin~\& Pariev~(1994, 1996)  
there exist so-called standing modes among
eigenmodes in jets ($v_{\rm group}=0$) which do not propagate with finite
velocity along the jet but are only subject to diffuse spreading and,
therefore, their amplitude is maximal.
The phase velocity of these modes is greater than the speed of light.
The crests of the wave move along the jet  with superluminal velocity causing
acceleration of the particles on the Alfv\'en resonance surface. These regions can be
the objects which are seen
as bright knots with typical sizes of the order of the wavelength of the
standing wave, which is about the radius of the jet.

\section{ Synchrotron emission. }\label{sec2}

\begin{figure}
\centering
\includegraphics[width=13cm]{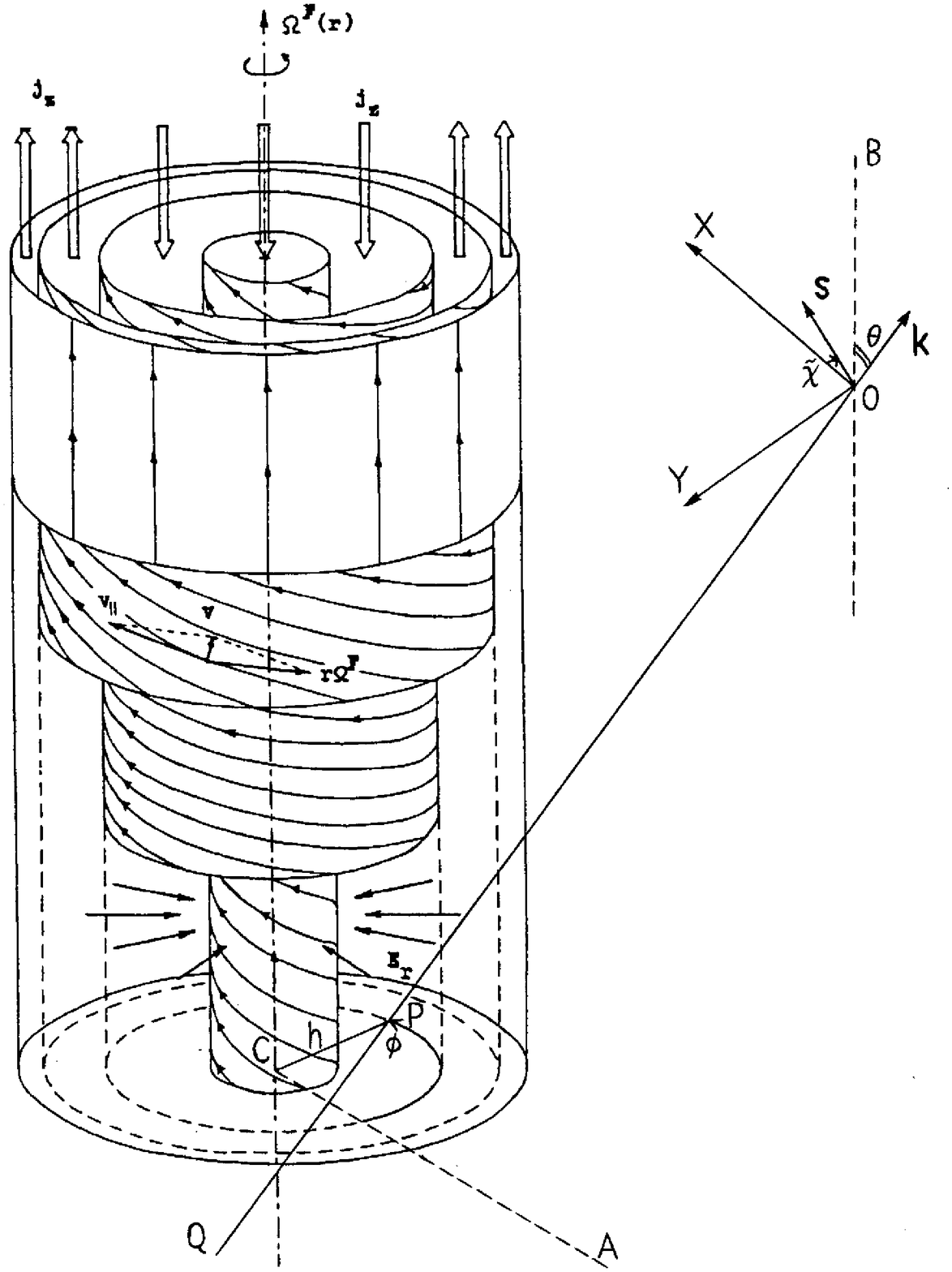}
\caption{The scheme illustrating the geometry of the jet and of the
polarization observations.
The jet boundary for $r=R$ and three magnetic tubes 
for $r=R/4$, $2R/3$ and $9R/10$ are shown for uniform $B_z$ and
angular velocity profile~(\protect\ref{eqnOmegaF}) with $\Omega=10$.
The magnetic field lines are spiralling on a magnetic tube.  
Since $\Omega^F(R)=0$, the total current 
through the jet is equal to zero and the magnetic field is purely poloidal 
both at the boundary and at the axis of symmetry. 
The density of the poloidal current $j_z$ is negative 
when $r<R/\sqrt{2}$ and positive when $R>r>R/\sqrt{2}$. 
The electric field ${\bf E}$ induced by jet rotation is radial. 
The plasma velocity ${\bf v}$ 
along the magnetic tube consists of two components: rotation with 
angular velocity $\Omega^F(r)$, and motion along the magnetic field lines 
with a speed ${\bf v}_\parallel=K{\bf B}$. 
A line of sight OQ passes through the jet at a closest distance 
$\mbox{CP}=h$ from the jet axis. The line CA is parallel to the plane
containing the observer and the jet axis. OXY is a rectangular coordinate
system on the plane of the sky of the observer, OX is in the plane
containing the observer and the jet axis. The line OB is parallel 
to the jet axis, and the wave vector of a photon ${\bf k}$ makes 
an angle $\theta$ with the line OB and the jet axis. Vector ${\bf s}$ is
a unit vector in the direction of the electric vector of 
the polarized component. Vector ${\bf s}$ is in the plane OXY and makes 
an angle ${\tilde\chi}$ with OX.
}
\label{fig1add}
\end{figure}

Let us now consider synchrotron emission of accelerated particles in a
cylindrical force--free jet with a spiral magnetic field.  
For cylindrical jet the stationary configuration of fields is (Istomin \&
Pariev 1994):
\beg 
{\bf B}=B_z{\bf e}_z+B_\phi{\bf e}_\phi;\quad B_\phi=\pm\frac{\omr}{c} B_z,
\label{eqn1}
\en
where $r,\phi,z$ are cylindrical coordinates. The profile of the poloidal 
magnetic field $B_z$ is taken to be uniform across the jet. 
The velocity of plasma can be written as
\beg 
{\bf u}=K{\bf B}+\omr{\bf e}_\phi,\label{eqn2} 
\en
where $\Omega^F$ is the angular rotational velocity of the magnetic 
field lines and $K{\bf B}$ is the component of the velocity parallel 
to the magnetic field. The electric field is induced due to the motion of 
highly conducting plasma across the magnetic field and is equal to
\beg
{\bf E}=-\frac{\omr}{c}[{\bf e}_\phi{\times\bf B}]. \label{eqn3}
\en
In components
\begin{eqnarray}
&& {\bf B}=(0,\pm\omr/c,1)B_z,\label{eqn3a} \\
&& {\bf E}=(-\omr/c,0,0)B_z, \label{eqn3b} \\
&& {\bf u}=(0, \omr(1\pm K B_z/c), KB_z).\label{eqn3c}
\end{eqnarray}
Here $\Omega^F$ and $K$ are functions of $r$, and $B_z$ does not depend on $r$.
The stationary electric field, which has an absolute value of $\omr B_z/c$, is
not small compared to the  magnetic field when $\omr \sim c$.
The rotation velocity $\omr$ can exceed the speed of light; 
nevertheless $u$ remains less than $c$ due to the existence of the 
predominantly toroidal magnetic field. This can be achieved by appropriate
choice of the parameter $K$. 
The natural requirement that the total poloidal current is closed inside the jet 
and that the total electric charge of the jet is zero implies that $\Omega^F$ 
must vanish at the boundary of the jet $r=R$ (Istomin~\& Pariev 1994).
Besides this requirement, the dependence $\Omega^F=\Omega^F(r)$ should be determined
by the conditions at the jet origin. 
An arbitrary parameter $K(r)$ in expression~(\ref{eqn2}) for the plasma velocity 
determines the radial profile of the longitudinal velocity.
$K(r)B_z$ is the velocity along the magnetic field which is not related
to the rotation. We choose a velocity ${\bf u}$ which minimizes the
kinetic energy of the plasma in the stationary reference frame
\beg
{\bf u}=\frac{(0,\omr,\mp (\omr)^2/c)
}{1+(\omr/c)^2}.\label{eqn4}
\en
It is probable that the plasma moves with that velocity in real jets
originating near black holes (Frolov~\& Novikov 1998). 
The geometry of the magnetic and electric fields and the flow in 
the jet are illustrated in Fig.~\ref{fig1add}. 

In accordance with our theory of particle acceleration inside a force--free 
relativistic jet (Beresnyak et al. 2002) 
we consider the distribution function of emitting particles to be isotropic
in momentum and power law in energy
\beg
dn=K_e\e^{-\g}d\e dVd\Omega_{\bf p} \label{eqn37} 
\en
Here $dn$ is the number of particles in the energy interval
$\e,\e+d\e$, $dV$ is the elementary volume,
$d\Omega_{\bf p}$ is the elementary solid angle in the direction of the
particle momentum ${\bf p}$, $K_e=K_e(r)$, $\g=\mbox{constant}$.
We will assume that ultrarelativistic particles are distributed according
to Eq.~(\ref{eqn37})
in the frame moving with the plasma with the velocity ${\bf u}$, Eq.~(\ref{eqn4}).
One needs to know the distribution of the density of relativistic particles
over the jet radius $K_e(r)$. Several mechanisms for particle 
acceleration inside jets have been considered previously. The possible 
processes are acceleration by the shocks in the supersonic flow of mass-loaded
jets (Blandford~\& Eichler 1987) and 
acceleration due to magnetic reconnection near the neutral
layers of the poloidal magnetic field (Romanova~\& Lovelace 1992). Usually,
the distribution of fast particles inside the jet was assumed to be 
uniform (Laing 1981) or smoothly varying with the radius in the jet 
(K\"onigl~\& Choudhuri 1985).    
According to our mechanism of particle acceleration (Beresnyak et al. 2002),
relativistic particles are accelerated close to the Alfv\'en resonance surface 
$r=r_A$ and are concentrated near the same surface. The position of 
the Alfv\'en resonance in a force--free jet is determined by the condition
\beg
kc-\omega+2m\Omega^F(r_A)-\left(\frac{\Omega^F(r_A) r_A}{c}\right)^2
(\omega+kc)=0, \label{eqnA}
\en
where $\omega=\omega_m(k)$ is the eigenfrequency of the perturbation
of the magnetic field $\propto \exp (-i\omega t + ikz+im\phi)$
with the wavenumber $k$ along the axis of the jet and azimuthal number
$m$.
We will consider two cases for the spatial distribution of particles, namely,
uniform in the jet volume and concentrated close to the Alfv\'en resonant
surface $r=r_A$. 

The influence of the plasma on the emission of relativistic $e^-$ and $e^+$ is
neglected in the present work. Absorption and reabsorption are also not 
taken into account. Electrons and positrons are accelerated in the 
same way near Alfv\'en resonance and have the same distribution function.
Therefore, the Faraday rotation effect is absent as well as intrinsic 
depolarization inside the jet. VLBI polarization measurements
are done at frequencies in the range $1\,\mbox{GHz}$ to $100\,\mbox{GHz}$.  
Synchrotron self-absorption
is very small at GHz frequencies and higher frequencies. 
For $B\sim 10^{-2}\,\mbox{G}$, 
$n\sim 0.1\,\mbox{cm}^{-3}$, and $\zeta\approx 2.5$ an order of magnitude estimate 
for the synchrotron self-absorption coefficient is $\displaystyle \kappa_{s}=
(5\cdot 10^3\,\mbox{pc})^{-1}\left(\nu/1\,\mbox{GHz}\right)^{-2-\zeta/2}$
(e.g., Zhelezniakov 1996). The corresponding optical depth through the 
$1\,\mbox{pc}$ jet is only $2\cdot 10^{-4}$. 
The typical energy of particles emitting in this frequency range in
$10^{-2}\,\mbox{G}$ magnetic field is $100\,\mbox{MeV}$ to $1\,\mbox{GeV}$.
Conversion of linear polarized radio waves into circular polarized 
(Cotton--Mouton effect) is also small. For both
thermal and relativistic power law particle distributions with 
$\zeta\approx 2.5$ and a minimum Lorentz factor $\sim 10$ the estimate for 
the conversion coefficient is $\displaystyle\kappa_{c}=(500\,\mbox{pc})^{-1}
\left(\nu/1\,\mbox{GHz}\right)^{-3}$ (Sazonov 1969). This is too small to 
influence the transfer of linear polarization inside the jet, but can be 
the source for a small, at the level of a fraction of per cent, circular 
polarization. We focus on linear polarization in this work.
 
At first, we neglect Doppler boosting, i.e. assume that
${\bf u}=0$ for the purpose of the calculation of the synchrotron radiation.
However, the magnetic field will still be given by formulae~(\ref{eqn1}) 
with nonvanishing $\Omega^F$. In Sect.~\ref{sec3} we take into account 
Doppler boosting with velocity ${\bf u}$ consistent with the field configuration
according to expressions~(\ref{eqn3a}--\ref{eqn3c}) and see that the proper
consideration of the Doppler effect leads to a marked change in polarization.  
When no Doppler effect is considered, 
the Stokes parameters are obtained by integration along the line of sight 
passing
through the region filled with relativistic particles inside the jet 
(e.g., Ginzburg 1989). 
Let us denote the angle between the axis of the jet
(the direction of the axial magnetic field $B_z$) and
the observer by $\theta$, such that $\theta=0$ when $B_z$ points exactly 
at the observer, $\theta=\pi/2$ when the line of sight of the observer 
is perpendicular to the direction of the jet axis, and $\theta=\pi$ when
the $B_z$ points exactly away from the observer. The azimuthal 
angle $\phi$ is measured from the plane containing  
the observer and the jet axis;  $\phi$ increases
from $0$ to $2\pi$ in anticlockwise direction if viewed from the 
positive direction of axial magnetic field $B_z$. 
Let us denote the distance between the line of sight 
and the jet axis by $h$ and the radius of the jet by $R$. 
We will take $h$ as a signed quantity: $h>0$ for lines of sight passing in 
the sector $0<\phi<\pi$, $h<0$ for the lines of sight passing in 
the sector $\pi<\phi<2\pi$. We will parametrize integration along the line 
of sight by integration in $\phi$ keeping the value of $h$ constant.
We define Stokes parameters in the  usual way with reference 
to the rectangular coordinates in the plane of
the sky with the $x$--axis oriented along the projection of the jet axis on 
the plane of the sky and the $y$--axis orthogonal to the 
$x$--axis in the anticlockwise
direction. This geometry and the angles are illustrated in Fig.~\ref{fig1add}.
Considering the geometry described above we obtain the following 
expressions for the Stokes parameters
\begin{eqnarray}
&& I=\frac{\g+7/3}{\g+1}k(\nu)\int\limits_{\phi_1}^{\phi_2} 
|B_\perp|^{(\g+1)/2}\frac{h}{\sin\theta\sin^2\phi}\,d\phi,\label{eqn38} \\
&& Q=k(\nu)\int\limits_{\phi_1}^{\phi_2}\frac{|B_\perp|^{(\g+1)/2}}{|B_\perp|^2}\frac{[B_\phi^2\cos^2\phi-
(B_z\sin\theta+B_\phi\sin\phi\cos\theta)^2]h}{\sin\theta\sin^2\phi}\,d\phi,\label{eqn39} \\
&& U=-2k(\nu)\int\limits_{\phi_1}^{\phi_2}\frac{|B_\perp|^{(\g+1)/2}}{|B_\perp|^2}B_\phi\cos\phi
(B_z\sin\theta+B_\phi\sin\phi\cos\theta)\frac{h}{\sin\theta\sin^2\phi}\,d\phi, \label{eqn39a} \\
&& V=0, \label{eqn39b}
\end{eqnarray}
where $B_\perp$ is the component of the magnetic field perpendicular to the line
of sight. The integration limits are
$\displaystyle \phi_1=\pi\,\mbox{sgn}\,h-\arcsin\frac{h}{R},
\quad\phi_2=\arcsin \frac{h}{R} $.
The value of $k(\nu)$ is a function of the observed frequency $\nu$ of the 
radiation
$$ k(\nu)=\frac{\sqrt3}4\Gamma\left(\frac{3\g-1}{12}\right)
 \Gamma\left(\frac{3\g+7}{12}\right)\frac{e^3}{m_e c^2}\left[
   \frac{3e}{2\pi m_e^3c^5}\right]^{(\g-1)/2}\nu^{-(\g-1)/2}\frac{K_e}{D^2},$$
where $D$ is the distance between the source and the observer, $e$ 
and $m_e$ are the charge and mass of an electron, and $\Gamma$ is 
the Euler gamma-function.

We denote the position angle of the electric field vector in the plane of 
the sky by $\chit$.
The angle $\chit$ is measured clockwise from the direction parallel
to the projection of the jet axis on the plane of the sky and $0<\chit<\pi$
(see Fig.~\ref{fig1add}).
For ultra-relativistic particles $V=0$, and the emission has only linear
polarization. The degree of polarization of the observed radiation is expressed
as $\Pi=\sqrt{Q^2+U^2}/I $, the resultant position angle of the electric field
$\chit_{\rm res}$ measured by the observer is found from
$$ \cos 2\chit_{\rm res}=\frac Q{\sqrt{Q^2+U^2}},
\quad \sin 2\chit_{\rm res}=\frac U{\sqrt{Q^2+U^2}},
\quad 0\le\chit_{\rm res}<\pi.$$
It is seen that when one replaces the variable $\phi$ by $\pi-\phi$ in 
expression~(\ref{eqn39a}), $U$
changes its sign and, therefore, $U=0$.
Consequently, if $Q>0$ then $\chit_{\rm res}=0$,
if $Q<0$ then $\chit_{\rm res}=\pi/2$. Thus, the observed electric field vector
is either parallel
or perpendicular to the jet axis on the plane of sky. This may account for the
bimodal distribution of the observed $\chit_{\rm res}$ (Gabuzda et al. 1994).
Now let us integrate expressions~(\ref{eqn38}) and~(\ref{eqn39}) for $I$ and $Q$  
over parameter $h$ in order to obtain the total radiation flux and polarization per
unit length of the jet image on the sky.  
Assuming uniform emissivity over the jet volume,
$B_z=\mbox{constant}$ and $B_\phi=B_z\Omega^F r/c$ expressions for
$I$ and $Q$ from the unit jet length become
\begin{eqnarray}
&& I=\frac{\g+7/3}{\g+1} k(\nu)\int\limits_{-R}^R dh
\int\limits_{\phi_1}^{\phi_2} |B_\perp|^{(\g+1)/2}
\frac{h}{\sin\theta\sin^2\phi}\,d\phi,\label{eqn40} \\
&& Q=k(\nu)\int\limits_{-R}^R dh\int\limits_{\phi_1}^{\phi_2}|B_\perp|^{(\g-3)/2}
\frac{[{\Omega^F}^2r^2\cos^2\phi-
(c\sin\theta+\Omega^F r\sin\phi\cos\theta)^2]B_z^2h}
{c^2\sin\theta\sin^2\phi}\,d\phi.\label{eqn41}
\end{eqnarray}

For $\g=3$ expressions~(\ref{eqn40}) and~(\ref{eqn41})
are reduced to the form
\begin{eqnarray}
&& I=\frac{4}{3}k(\nu)B_z^{2}
\frac\pi{\sin\theta}\int\limits_0^R2r
\left[\sin^2\theta+
\frac{{\Omega^F}^2r^2}{c^2}\left(1-\frac12\sin^2\theta\right)
\right]\,dr,\label{eqn41a} \\
&& Q=k(\nu)B_z^{2}\pi\sin\theta\int\limits_0^R\left(\frac{{\Omega^F}^2r^2}{c^2}-2
\right)r\,dr. 
\label{eqn41b}
\end{eqnarray}
From the above we see that in the case $\g=3$ the sign of $Q$ does not depend
on $\theta$ since $\sin\theta>0$ for any $0<\theta<\pi$. 
The sign of $Q$ is determined by the function $\Omega^F(r)$ only.

Let us choose a particular dependence of $\Omega^F(r)$ satisfying the 
requirements $\Omega^F(R)=0$ and $d\Omega^F/dr=0$ at $r=0$. At present, there 
is no well established theory which would predict $\Omega^F(r)$. 
Therefore, we choose the simplest monotonic profile satisfying the physical
requirements above: 
\beg
\Omega^F=\Omega \frac{c}{R}\left(1-\left(\frac{r}{R}\right)^2\right),
\label{eqnOmegaF}
\en
where $\Omega$ is a dimensionless parameter.
Although the exact position of $r_A$,
the ratio of $B_{\phi}/B_z$ at $r_A$ and the velocity of the flow in 
the vicinity of $r_A$ (properties which determine the observed 
polarization) are all dependent on the particular choice of
$\Omega^F(r)$ profile, it is the magnitude of 
the angular rotational velocity which influences all these parameters
most significantly. A qualitative difference may appear if 
the profile of $\Omega^F$ is non-monotonic, such that many Alfv\'en resonant
surfaces will exist inside the jet. In this work we limit our consideration
to variation of parameter $\Omega$ only, but keeping the radial profile 
of $\Omega^F$ monotonic and given by expression~(\ref{eqnOmegaF}).
Then, for $\g=3$ using expressions~(\ref{eqn41a}) and~(\ref{eqn41b})
one obtains the degree of polarization $\Pi$ 
$$ \Pi=\frac{3}{4}\frac{\displaystyle\left(\frac{\Omega^2}{24}-1\right)\sin^2\theta}
{\displaystyle
\sin^2\theta+\frac{\Omega^2}{12}\left(1-\frac{1}{2}\sin^2\theta\right)}. $$
If $|\Omega|>2\sqrt6$, then $Q>0$ and $\chit_{\rm res}=0$, i.e. the electric vector 
of the polarized emission is parallel to the projection of the jet axis on the plane of
the sky. This does not mean, however, that the orientation of the 
magnetic field in the jet is purely perpendicular to the jet axis. Clearly, one can
see from Eq.~(\ref{eqn1}) that the poloidal magnetic field is comparable to 
the toroidal field.   
If $0<|\Omega|<2\sqrt6$ then $Q<0$ and $\chit_{\rm res}=\pi/2$, i.e. the electric vector 
of the polarized emission is perpendicular to the projection of the jet axis on the plane of
sky. Again, this does not mean that the orientation of the 
magnetic field in the jet is purely parallel to the jet axis. The toroidal magnetic field
vanishes only if $\Omega=0$, otherwise the toroidal magnetic field is comparable to 
the poloidal magnetic field. This simplified analytical calculation shows that one needs 
to be very careful when interpreting the polarization measurements in terms of the
orientation of the magnetic field.

\begin{figure}
\centering
\includegraphics[width=0.45\textwidth]{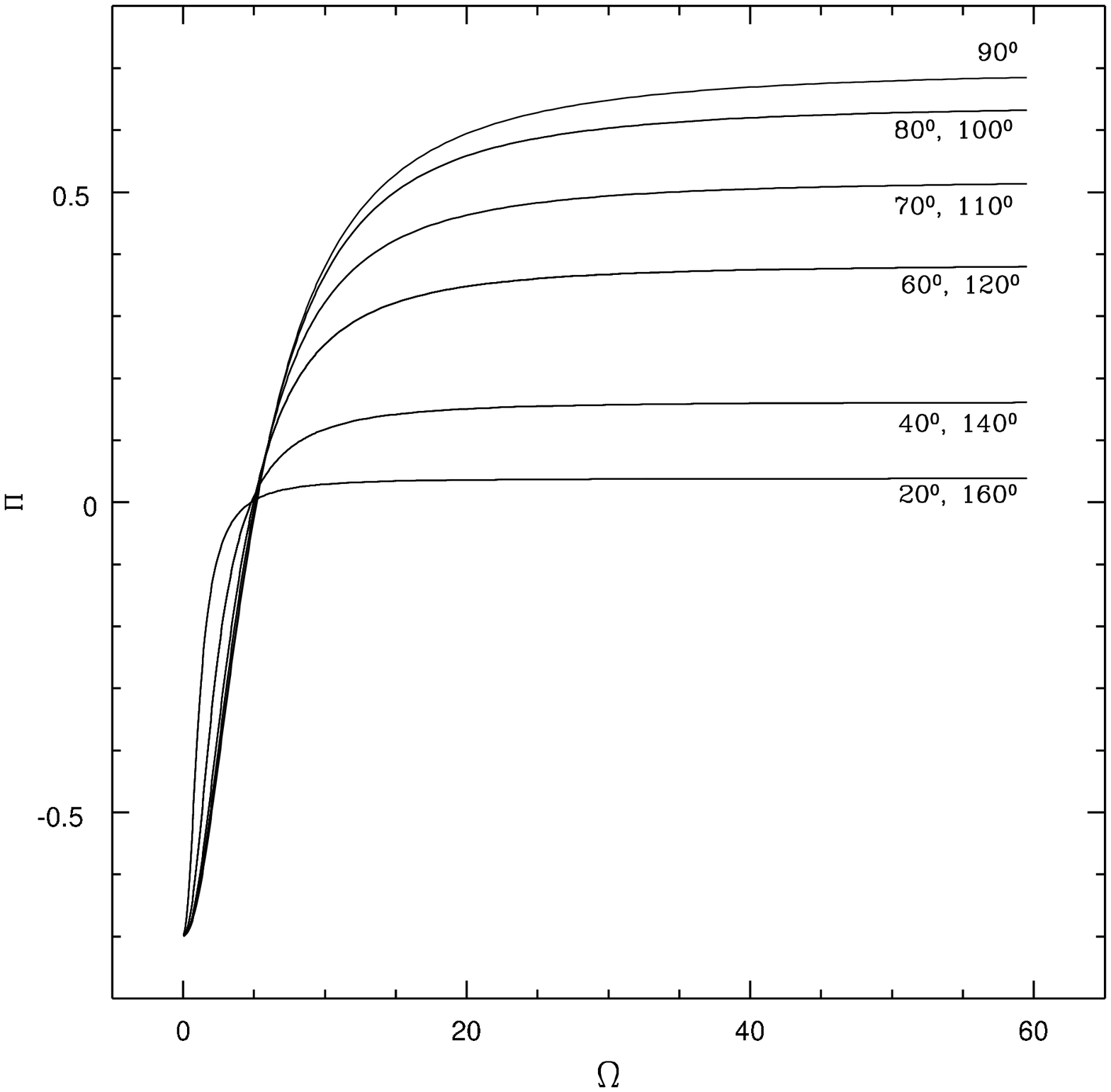}
\hfill
\includegraphics[width=0.45\textwidth]{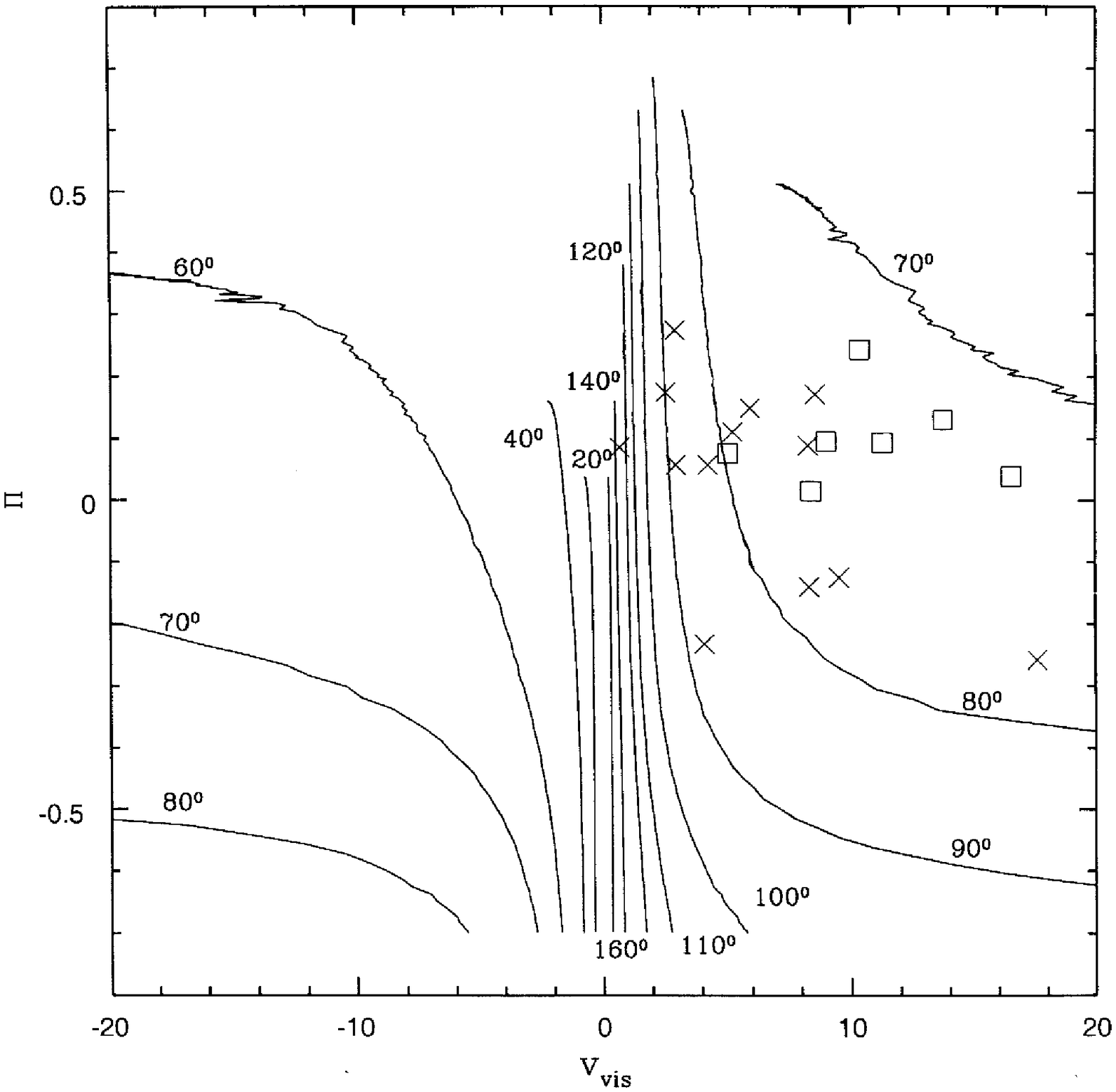}
\caption{Dependences of linear polarization $\Pi$ on angular
rotational velocity $\Omega$ (left panel) and on visual velocity of
knots~$v_{\rm vis}$ (right panel) for fundamental axisymmetric mode $m=0$.
$\Pi>0$ when the electric field
vector is oriented parallel
to the jet axis, $\Pi<0$ when the electric field vector is oriented
perpendicular to the jet axis.
$v_{\rm vis}>0$ means that the observed motion of
knots is direct, $v_{\rm vis}<0$ means reversal motion.
Crosses are for the BL~Lac objects and
squares are for the quasars listed in Table~1.}
\label{fig1}
\end{figure}

Eqs.~(\ref{eqn40}) and~(\ref{eqn41}) were integrated
numerically for the dependence $\Omega^F(r)$ given by expression~(\ref{eqnOmegaF})
and $\g=2.1$, which corresponds to the power law
index in the spectrum of synchrotron emission of $(\g-1)/2=0.55$, the mean
observed value for extragalactic jets (Scheuer 1984).
The results are presented in Fig.~\ref{fig1}. Positive values of the degree of
polarization
correspond to $\chit_{\rm res}=0$, negative to  $\chit_{\rm res}=\pi/2$. In the case
$\Pi=0$ the radiation is entirely unpolarized. Note also that in the nonrelativistic
consideration $\Pi(180^0-\theta)=\Pi(\theta)$ for any value of $\theta$.

Istomin~\& Pariev (1996) presented a hypothesis that knots observed
in jets can be the manifestation of the  ``standing wave'' phenomenon.
If one knows the dispersion curves $\om=\om(k)$ and wavenumbers~$k_{\rm min}$ at
which $d\om/dk=0$
for perturbations, one can calculate the velocity of moving crests of the
``standing wave'' as
$ v_{\rm phas}=\om_{\rm min}(k_{\rm min})/k_{\rm min}$
and corresponding observed velocity $\displaystyle v_{\rm vis}=\frac{v_{\rm phas}\sin\theta}
{1-v_{\rm phas}\cos\theta/c}$.
Note also that changing the sign of~$B_{\phi}$ in Eq.~(\ref{eqn3a}) is equivalent to
changing the sign of $\Omega^F$, since $I$ and $Q$ depend only on $B_z$ and
$B_{\phi}$ and not on $u$ or $E_r$.
In Fig.~\ref{fig1} we present the relations
of the polarization $\Pi$ of the jet emission and $v_{\rm vis}$ obtained in the case
of $\Omega$ changing from $0$ to $100$ for some angles $\theta$.
We also plotted observational points for jets in
BL Lac objects and quasars for which
we were able to find both polarization and proper motion measurements in
the literature and which are listed in Table~1.
For each object we averaged the degree
of polarization and observed velocity over all bright knots
observed in a particular object excluding a few knots, for which
peculiarity was indicated in the original references. This procedure might be very questionable
although because of the lack of observations of the emission from the space between
knots on parsec scale we had nothing better. The polarization measurements also
depend on the frequency of observations. Specifically, different observations of
quasars give very different values of the polarization of the knots at centimetre and
millimetre wavelengths.
Numbers for $v_{\rm vis}$ were
calculated assuming a Hubble constant $H=65 h\mbox{km/s}\cdot\mbox{Mpc}$
\begin{table}
\caption{Polarization degrees and velocities of knots in BL~Lac objects
and quasars}

\begin{tabular}{llccc}
Class & Source & $\Pi\,$,\% & $v_{\rm vis}h/c$ & References \\
\hline
BL~Lac & 0138-097 & 8.8 & 8.5 & 1 \\
BL~Lac & 0454+844 & 8.5 & 0.92 & 1 \\
BL~Lac & 0735+178 & -14. & 8.5 & 1 \\
BL~Lac & 0828+493 & -12.5 & 9.7 & 1 \\
BL~Lac & 0851+202 & -23.3 & 4.3 & 1 \\
BL~Lac & 0954+658 & 17.2 & 8.8 & 1 \\
BL~Lac & 1418+546 & 5.6 & 3.2 & 1 \\
BL~Lac & 1732+389 & -25.9 & 17.8 & 1 \\
BL~Lac & 1803+784 & 27.5 & 3.2 & 2,3 \\
BL~Lac & 1823+568 & 14.9& 6.2& 1 \\
BL~Lac & 2007+777 & 5.6 & 4.5 & 1 \\
BL~Lac & 2200+420 & 11. & 5.5 & 1 \\
quasar & 0212+735 & 1.4 & 8.6 & 2,3 \\
quasar & 0836+710 & 2.2 & 21.7 & 2,3 \\
quasar & 0906+430 & 7.5 & 5.3 & 2,3 \\
quasar & 1055+018 & 9.5 & 9.23 & 4  \\
quasar & 1308+326 & 3.8 & 16.8 & 5 \\
quasar & 3C345    & 11. & 29.3 & 2,3 \\
quasar & 1642+690 & 13. & 14.0 & 2,3 \\
quasar & 1928+738 & 9.3 & 11.5 & 2,3 \\
quasar & 3C380    & 24.4 & 10.6 & 2,3 \\
\end{tabular}

\medskip
RefeRences.--- (1)~Gabuzda et al. (2000),
(2)~Lister et al. (2001), (3)~Lister (2001),
(4)~Lister et al. (1998),
(5)~Gabuzda \& Cawthorne (1993)
\end{table}

In the case of accelerated particles concentrated very close to the Alfv\'en
surface $r=r_A$ expressions~(\ref{eqn40}) and (\ref{eqn41}) can be reduced to integrals in
$\phi$ only, while $r=r_A$.
\begin{figure}
\centering
\includegraphics[width=0.45\textwidth]{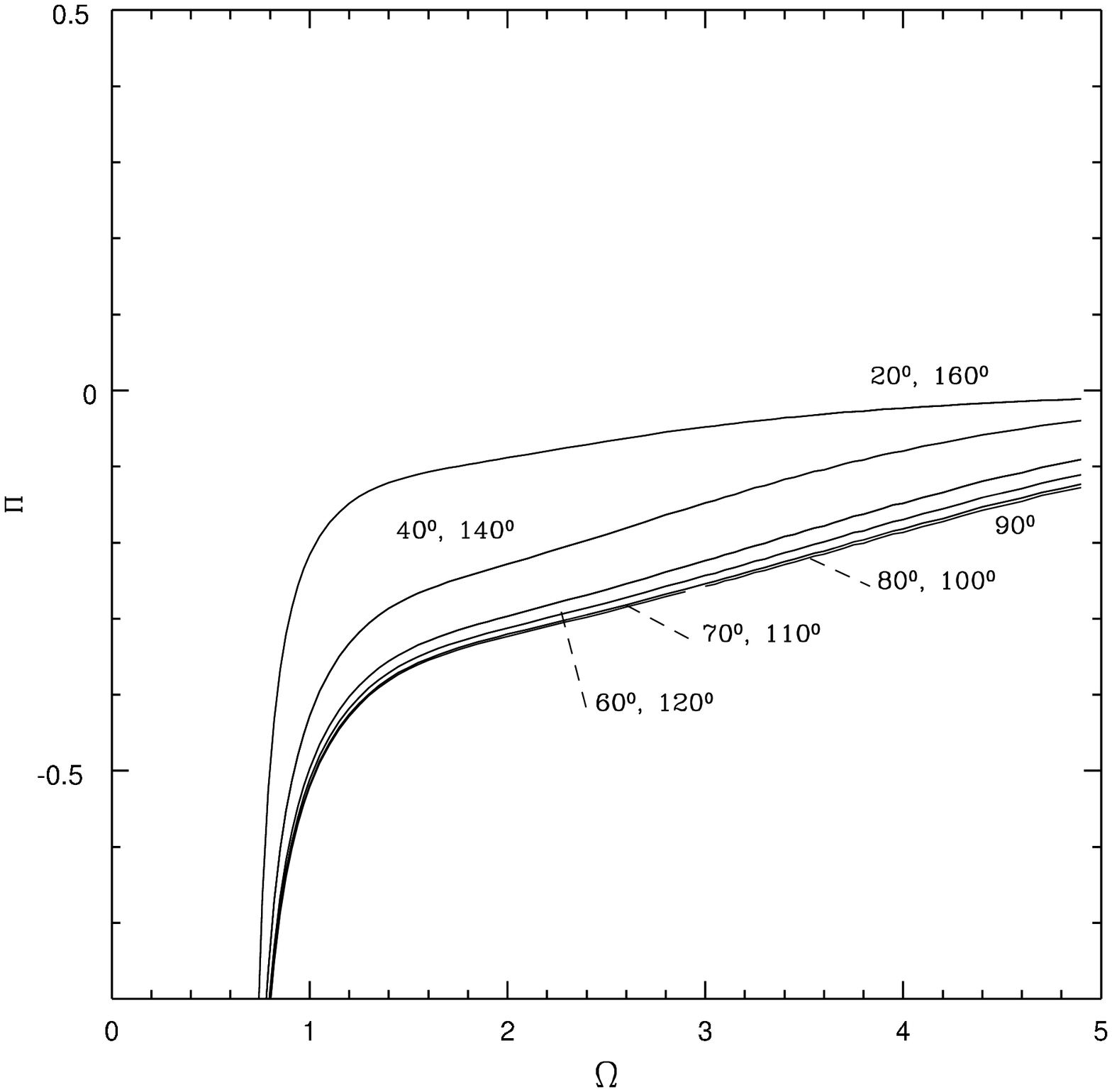}
\hfill
\includegraphics[width=0.45\textwidth]{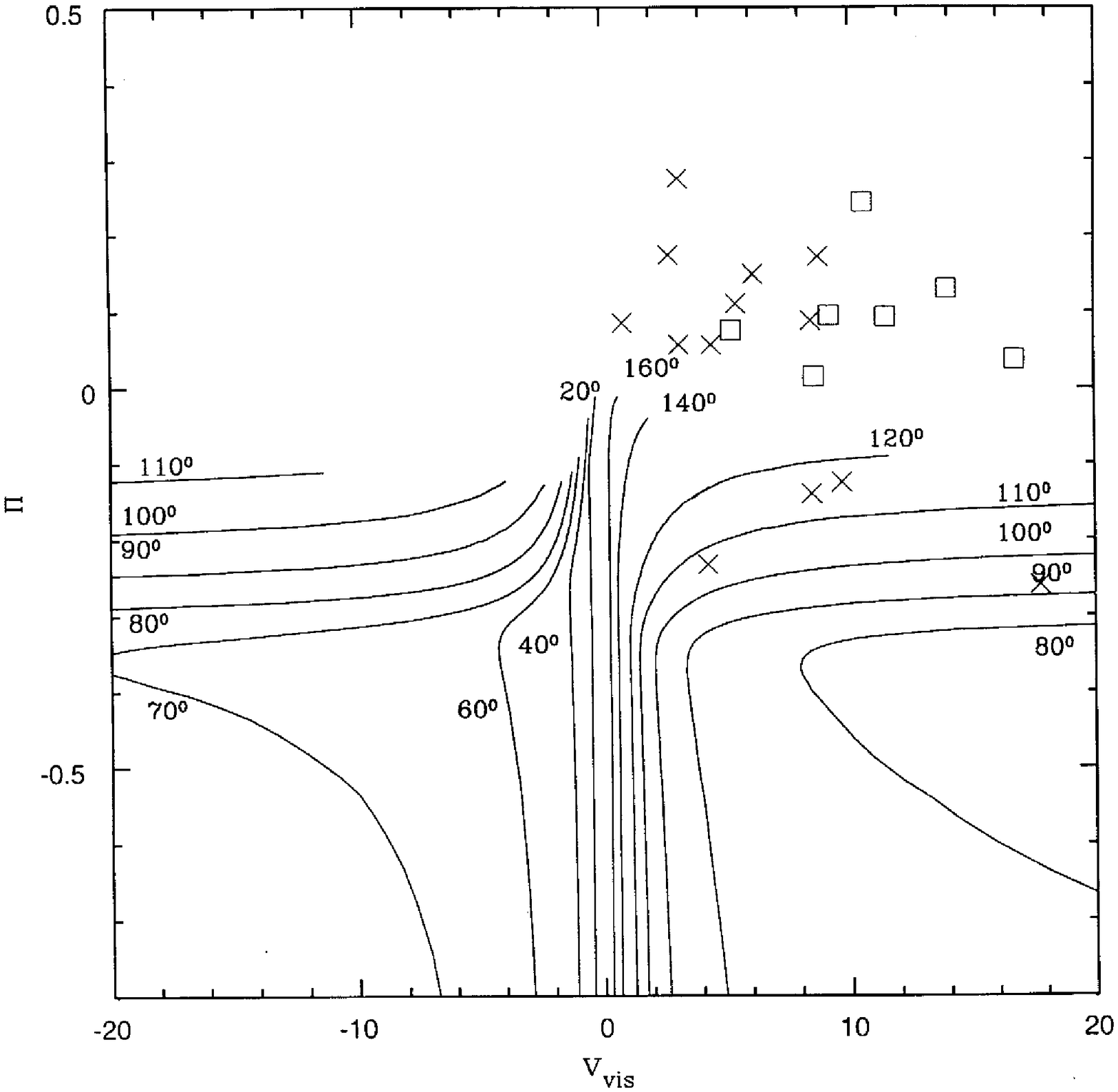}
\caption{The same as in Fig.~\protect\ref{fig1} 
but emitting particles are concentrated
close to the Alfv\'en resonance surface, $m=1$ fundamental mode.
The curves are plotted only for $0.7<\Omega<4.9$, when the Alfv\'en resonance
exists for the $m=1$, $k=k_{\rm min}$ mode.
}
\label{fig2}
\end{figure}
Since for axisymmetric modes ($m=0$) always
$|\omega_{\rm min}|>|k_{\rm min}|c$~(Istomin~\& Pariev 1994) one can see
from the expression~(\ref{eqnA})
that the Alfv\'en resonance does not exist for axisymmetric modes.
Therefore, there should be no particle acceleration for this mode. The next
most important modes are $m=1$ and $m=-1$. 
Since none of the quantities $k_{\rm min}$,
$\omega_{\rm min}$ or $\Pi$ changes when~$m$ changes to~$-m$ and $\Omega^F$ to
$-\Omega^F$, it is enough to consider only positive $m=1$ and 
all values of $\Omega^F$ (positive and negative). 
Polarization $\Pi$ versus $v_{\rm phas}$ for this case is plotted
in Fig.~\ref{fig2}
as well as observational points. We plotted $v_{\rm phas}$ only for the 
first non-axisymmetric
fundamental mode $m=1$ which is expected to be excited more easily than
higher modes with $m>1$. Observations of the parsec scale jet in BL~Lac 
object 0820+225 (Gabuzda et al. 2001) 
show the S--like shape which can be attributed to the mode $m=1$. 
The polarization curves are plotted only for values
of $\Omega$ at which the Alfv\'en resonance $r=r_A$ is present inside the
jet. As previously, $\Omega^F(r)$ is given by expression~(\ref{eqnOmegaF})
and $\g=2.1$.
It is seen that in the case of uniform particle distribution across the
jet (Fig.~\ref{fig1}), curves for $\theta$ changing 
from $120^\circ$ to $70^\circ$  are
located in the region where points from observations of
BL~Lac object (crosses) and quasars (squares) are. However,
the polarization curves for emitting particles concentrated on the Alfv\'en
resonance surface (Fig.~\ref{fig2}), 
as is the case for our acceleration mechanism,
do not match the observations of quasars. It is seen from Fig.~\ref{fig2} that the 
electric field vector is always oriented perpendicular to the jet axis unlike 
the case with uniform distribution of relativistic particles across the jet.
The reason for this difference is that the Alfv\'en resonance surface exists 
only for moderate values of $\Omega$, when the toroidal magnetic field is not 
strong enough to cause the $\chit_{\rm res}$ to become oriented parallel to 
the projection of the jet axis on the plane of the sky.

\section{Relativistic Doppler boosting effect on polarization}\label{sec3}

Now let us make more realistic calculations taking into account
Doppler boosting and the presence of a strong electric field. As was first
pointed out by Blandford \& K\"onigl~(1979) the effect of relativistic
aberration of light causes a swing of the direction of polarization of
synchrotron emission when the source is Doppler boosted. This is due
to the fact that the electric vector of the polarized component of the
wave must remain orthogonal to the wave vector.
We will denote values measured in the plasma rest frame by primes. 
Let us find the transformation rules for the polarization tensor
$J_{\alpha\beta}=<E_\alpha E_\beta^*>$ from the local reference 
frame of the small emitting volume moving with velocity ${\bf u}$ 
to the observer reference frame. Here $<\ldots >$ is the averaging 
over realizations of electric wave fields. The Stokes parameters are components 
of the polarization tensor
$J_{\alpha\beta}=\frac12\left(\begin{array}{cc}
                              I+Q & U\\
                              U & I-Q
                              \end{array}\right).$
We denote the local Cartesian coordinates in the source frame by $x' y' z'$ 
and the Cartesian coordinates in the observer frame
by $xyz$. The transformation can be done by
directly using the transformation formulae for the electric and
magnetic fields of the wave (Cocke \& Holm 1972). The result is
\beg
J_{\alpha\beta}=J^\prime_{\alpha'\beta'}
\frac{1-u^2/c^2}{(1-u/c\cos\delta)^2}. \label{eqn42}
\en
Here we denote by $\delta$ the angle between velocity ${\bf u}$ of the
source and the direction toward observer~${\bf k}$ measured in the observer
frame; correspondingly, $\delta'$ is the same angle but measured in the
source frame.
Thus, the polarization ellipse in the observer frame is obtained
from the polarization ellipse in the source frame by rotation by the angle
$\delta-\delta'$ in the plane containing the direction to the
observer ${\bf k}$
and ${\bf u}$. Therefore, the degree of polarization is not changed, which
is obvious.
The Stokes parameters for the radiation of one particle 
are expressed through the radiation flux with two
main directions of polarization $p^{(1)}_{\nu'}$ and
$p^{(2)}_{\nu'}$ by the formulae $I'=p^{(1)}_{\nu'}+p^{(2)}_{\nu'}$,
$Q'=(p^{(1)}_{\nu'}-p^{(2)}_{\nu'})\cos2\chit'$ ,
$U'=(p^{(1)}_{\nu'}-p^{(2)}_{\nu'})\sin2\chit'$,  
where $\chit'$ is the position angle of the electric vector of polarized 
radiation in the reference frame comoving with the emitting volume
(Ginzburg 1989).
Then, $p^{(1)}_{\nu'}$ and
$p^{(2)}_{\nu'}$ are transformed to the observer frame as
$$p^{(1)}_{\nu}=p^{(1)}_{\nu'}\frac{d\nu'}{d\nu}\frac{1-u^2/c^2}{(1-u/c\cos\delta)^2},
\quad p^{(2)}_{\nu}=p^{(2)}_{\nu'}\frac{d\nu'}{d\nu}\frac{1-u^2/c^2}{(1-u/c\cos\delta)^2}.$$
Therefore the intensity in the observer frame integrated along the line of sight is
$$ I(\nu,{\bf k})=\int(p^{(1)}_{\nu'}(\nu',\e',{\bf L},\chi',\psi')+
                       p^{(2)}_{\nu'}(\nu',\e',{\bf L},\chi',\psi'))\times $$
$$
\frac{d\nu'}{d\nu}\frac{1-u^2/c^2}{(1-u/c\cos\delta)^2}K_e{\e'}^{-\g}
\frac{dV'}{dV}d\e'd\Omega_{{\bf p}'}dL, $$
where $dL$ is the element of length along the line of sight, 
$\chi'$ is the angle between the momentum of the particle and the
magnetic field vector, $\psi'$ is the angle between the direction of
radiation and the cone formed by the velocity vector of the particle
rotating around the magnetic field line, both measured
in the frame comoving with the fluid element.
We will integrate over the momenta in the plasma rest frame while
integrating over the space coordinates in the observer frame.
Integration over the solid angle $\Omega_{{\bf p}'}$ in the momenta space
in a comoving frame may be done analogous to the usual case of 
synchrotron emission of ultra-relativistic particles,
when there is no electric field ${\bf E=-u\times B}$. Integration over
$\Omega_{{\bf p}'}$ and $\e'$ leads to (Ginzburg 1989)
\beg
I(\nu,{\bf k})=\frac{\g+7/3}{\g+1}k(\nu)
\int \left(\frac{1-u/c\cos\delta}{\sqrt{1-u^2/c^2}}
\right)^{-\frac{\g-1}{2}}(B'\sin\chi')^{\frac{\g+1}{2}}
\frac{dL}{1-u/c\cos\delta}. \label{eqn43} 
\en
Here $\displaystyle\nu'=\nu\frac{1-u/c\cos\delta}{\sqrt{1-u^2/c^2}}$,
${\bf B'}$ is the magnetic field in the emitter frame. The expressions
for $Q$ and $U$ can be obtained in a similar way. The result is
\begin{eqnarray}
&& Q(\nu,{\bf k})=k(\nu)
\int {\left(\frac{1-u/c\cos\delta}{\sqrt{1-u^2/c^2}}\right)}^{-\frac{\g-1}{2}}(B'\sin\chi')^{\frac{\g+1}{2}}
\frac{\cos 2\chit\, dL}{1-u/c\cos\delta},\label{eqn43a}  \\
&& U(\nu,{\bf k})=k(\nu)\int {\left(\frac{1-u/c\cos\delta}
{\sqrt{1-u^2/c^2}}\right)}^{-\frac{\g-1}{2}}(B'\sin\chi')^{\frac{\g+1}{2}}
\frac{\sin 2\chit\, dL}{1-u/c\cos\delta}. \label{eqn43b}
\end{eqnarray}
The component of the magnetic field perpendicular to the momentum of the particle in 
the comoving frame,
$B'\sin\chi'$, is obtained by applying the transformation rules 
\begin{eqnarray}
&& B'^2\sin^2\chi'=B^2_\phi+B_z^2-E_r^2- \nonumber\\
&& \frac{[-E_r(c^2-u^2)(u_z\sin\phi\sin\theta+u_\phi\cos\theta)+
u(c\cos\delta-u)(u_\phi B_\phi+u_zB_z)]^2}
{u^4(c-u\cos\delta)^2}. \label{eqn44} 
\end{eqnarray}
The expression for the position angle of the electric vector of the polarized radiation 
from a jet element in the observer frame $\chit$ is obtained from $\chit'$ by 
projecting the polarization ellipse in the plane perpendicular to ${\bf k}'$ 
on the plane perpendicular to ${\bf k}$. After some geometrical considerations the 
result for $\chit$ can be expressed as follows
\beg
\chit=\chit_u+\sigma,\label{eqn45}
\en
where angle $\chit_u$ can be found from 
\begin{eqnarray}
&& \tan\chit_u=\frac{E_r(u-c\cos\delta)(u_z\sin\phi\sin\theta+u_\phi\cos\theta)-
   u\sin^2\delta(u_\phi B_\phi+u_z B_z)}
  {E_ru\sin\theta\cos\phi(c-u\cos\delta)},\label{eqn45a}  \\
&& 0\le\chit_u<\pi,\nonumber
\end{eqnarray}
and angle $\sigma$, $0<\sigma<2\pi$, can be found from
\beg
\cos\sigma=-\frac{u_\phi\sin\phi\cos\theta+u_z\sin\theta}{u\sin\delta},
\quad \sin\sigma=-\frac{u_\phi\cos\phi}{u\sin\delta},\label{eqn45b} 
\en
where the expression for $\delta$ is
\beg
u\cos\delta=-u_\phi \sin\theta\sin\phi+u_z \cos\theta, \quad 0<\delta<\pi.
\label{eqn45c} 
\en 
Note, that by simply putting ${\bf u}=0$ in Eqs. 
(\ref{eqn43}--\ref{eqn45c}) one 
recovers the formulae for the emission from a jet with $B_{\phi}=0$
and $E_r=0$ . In order to recover Eqs.~(\ref{eqn40}--\ref{eqn41}) 
one needs to keep $B_{\phi}$ and $E_r$ unchanged but make 
no distinction between the
reference frame of the emitting element in the jet and the observer
frame.
For uniform particle distribution the intensities from a unit
length of the jet are
\begin{eqnarray}
&& I=\frac{\g+7/3}{\g+1}\frac{k(\nu)}{\sin\theta}\int_{0}^R r\,dr\int_0^{2\pi}\,d\phi
\left(\frac{1-u/c\cos\delta}{\sqrt{1-u^2/c^2}}\right)^{-\frac{\g-1}{2}}
\frac{|B'\sin\chi'|^{(\g+1)/2}}{1-u/c\cos\delta}, \nonumber \\
&& Q=\frac{k(\nu)}{\sin\theta}\int_{0}^R r\,dr\int_0^{2\pi}\,d\phi
\left(\frac{1-u/c\cos\delta}{\sqrt{1-u^2/c^2}}\right)^{-\frac{\g-1}{2}}
|B'\sin\chi'|^{(\g+1)/2}\frac{\cos2\chit}{1-u/c\cos\delta}\mbox{,}\label{eqn46} \\
&& U=\frac{k(\nu)}{\sin\theta}\int_{0}^R r\,dr\int_0^{2\pi}\,d\phi
\left(\frac{1-u/c\cos\delta}{\sqrt{1-u^2/c^2}}\right)^{-\frac{\g-1}{2}}
|B'\sin\chi'|^{(\g+1)/2}\frac{\sin2\chit}{1-u/c\cos\delta}\nonumber \mbox{.}
\end{eqnarray}
When the particles are located close to the Alfv\'en resonance
surface in the region $|r-r_A|\le\Delta$ and $\Delta\ll R$, the integration
in formulae~(\ref{eqn46}) reduces to integration in $\phi$ only while $r=r_A$.
It can be checked that under the change of $\phi$ to $-\phi$ in 
Eqs.~(\ref{eqn46}) the value of $Q$ is not changed,
and the sign of $U$ is reversed.
Thus, as a result of integration, $U=0$ and either $\chit_{\rm res}=0$ or
$\chit_{\rm res}=\pi/2$. This result of the bimodality of the values 
of $\chit_{\rm res}$ is the same as in Sect.~\ref{sec2} where we do 
not take into account Doppler boosting and electric field.

We perform numerical integration of Eqs.~(\ref{eqn46}) using 
formulae~(\ref{eqn44})--(\ref{eqn45c}).
\begin{figure}
\centering
\includegraphics[width=0.3\textwidth]{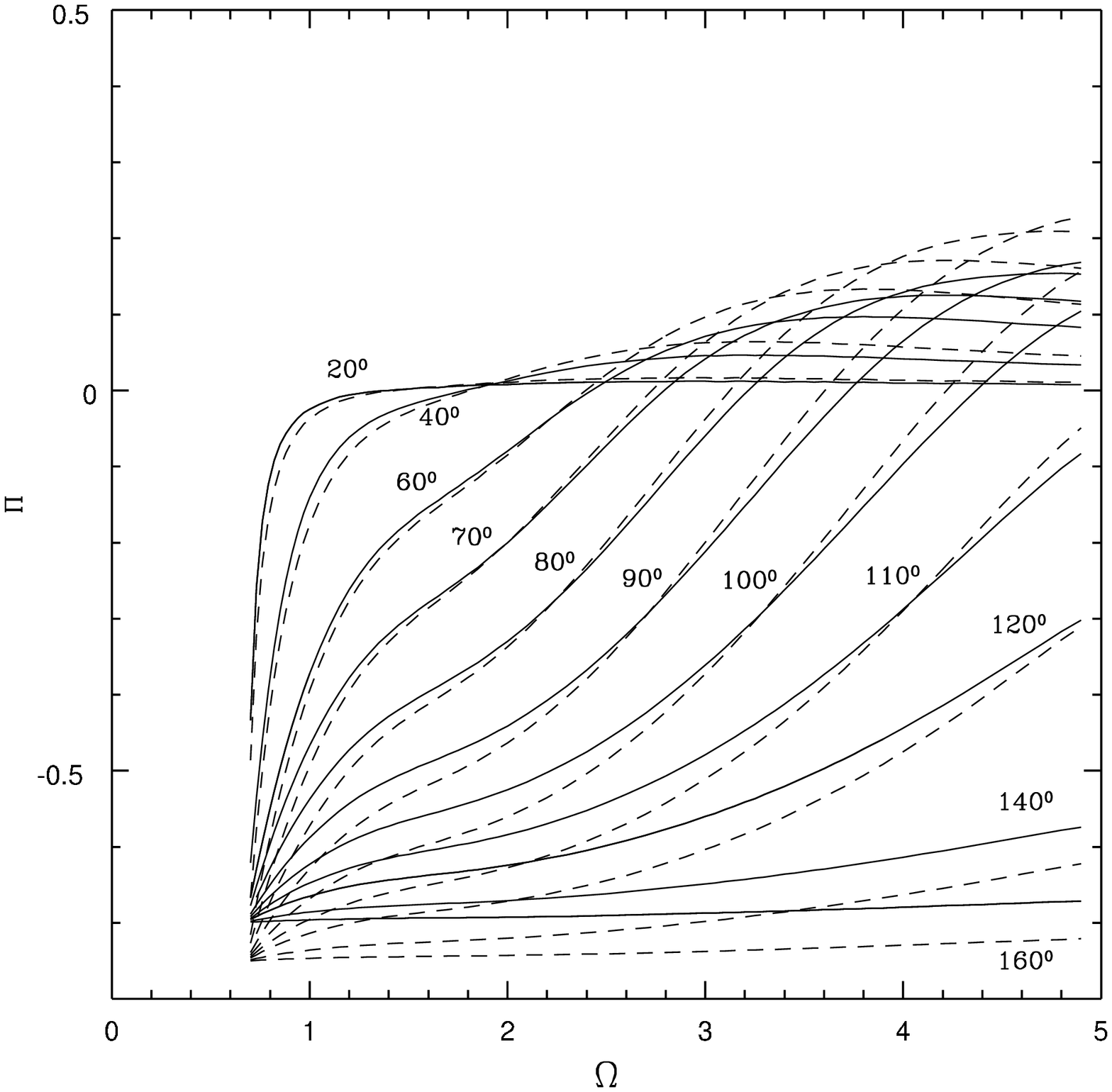}
\hfill
\includegraphics[width=0.3\textwidth]{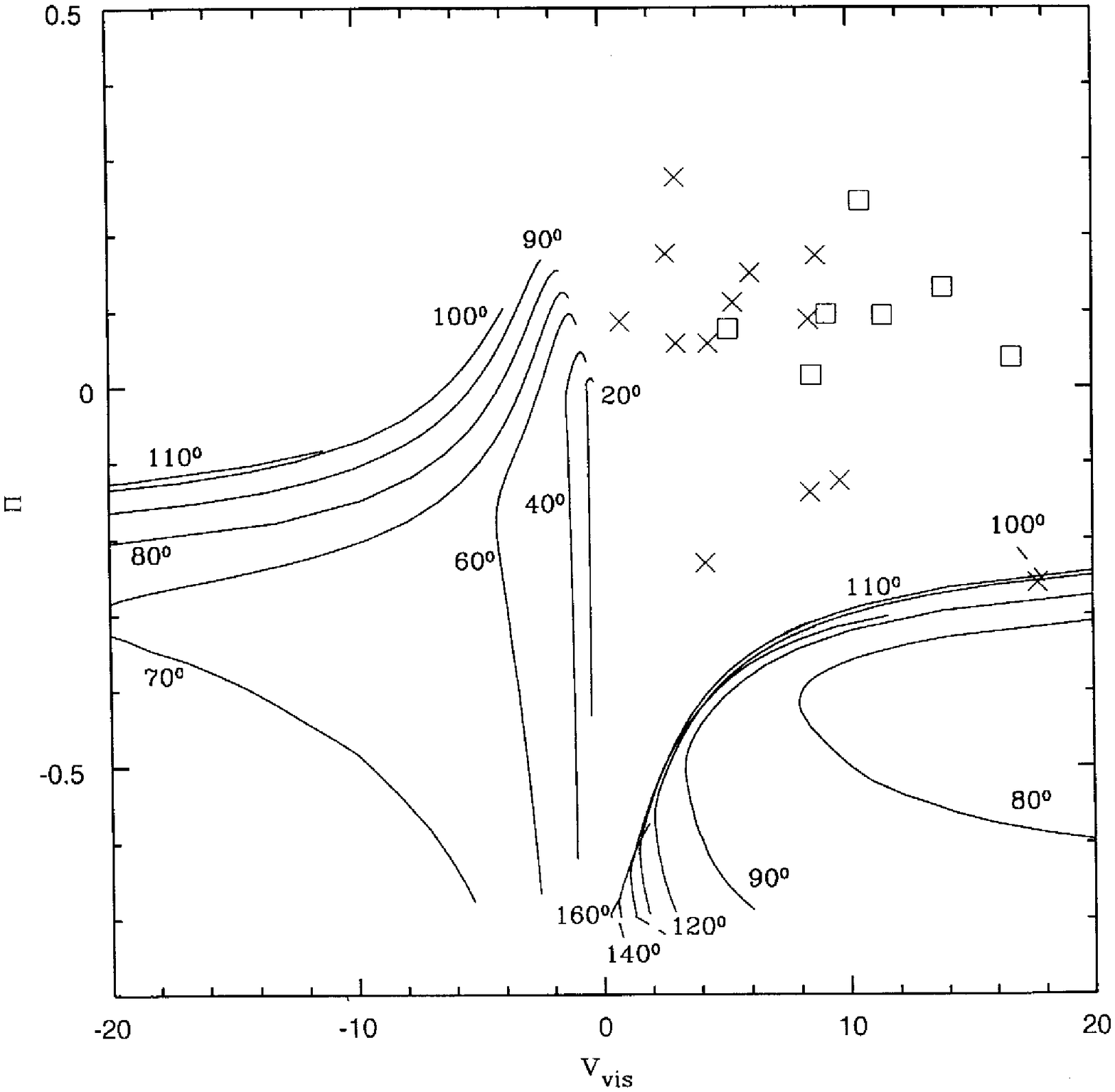}
\hfill
\includegraphics[width=0.3\textwidth]{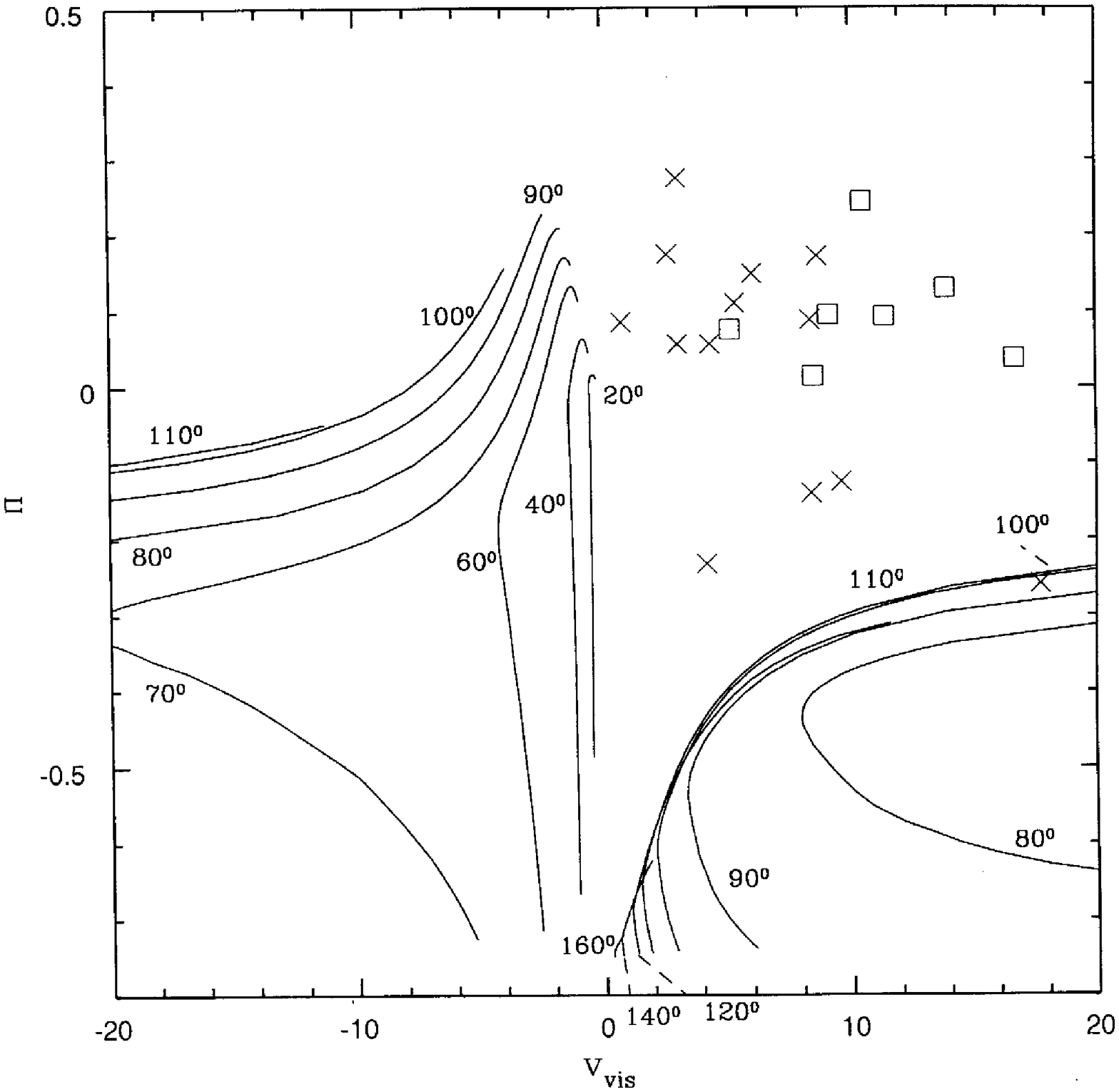}
\caption{Dependence of the polarization $\Pi$ on $\Omega$ (left panel), 
on $v_{\rm vis}$ (middle panel)
for $\g=2.1$ and for $\g=3$ (right panel). The emitting particles are
concentrated close to the Alfv\'en resonant surface.
Here $B_\phi=+\Omega^F r B_z/c$, fundamental eigenmode with $m=1$.
Notations are the same as in Fig.~\protect\ref{fig1}. In the left panel
solid curves are for $\g=2.1$, dashed curves are for $\g=3.$}
\label{fig3}
\end{figure}
As in Sect.~\ref{sec2} we take $B_z=\mbox{constant}$,
$B_\phi=\pm \Omega^F r B_z/c$, and take the dependence $\Omega^F(r)$ given 
by expression~(\ref{eqnOmegaF}).

The results for uniform particle distribution do
not match the observations. Taking into account the presence of the electric
field and Doppler boosting changes the polarization of the synchrotron
emission essentially. 
The apparent direction of the magnetic field, as it is usually
derived from polarization measurements (just perpendicular to the electric
vector in the polarized component of the radiation), is now transversal to the
jet axis only for small angle $\theta$ and large~$\Omega$. 
For most inclination angles~$\theta$ and angular velocities
$\Omega$ the apparent magnetic field is longitudinal (Fig.~\ref{fig3}). 
One can also see from Fig.~\ref{fig3}, where we plot~$\Pi$
versus~$v_{\rm vis}$ for $B_\phi=+\Omega^F r B_z/c$ and 
for the particles concentrated near Alfv\'en resonance surface,
that our simulation does not match the observational data from Table~1. 
However, for $B_\phi=-\Omega^F r B_z/c$ we are able 
to reproduce the observations well enough,
if the particles are distributed close to the Alfv\'en surface and $m=1$ or
$m=-1$. In both cases, with and without considering the Doppler boosting effect and
the electric field, $\Pi$ is not changed when one changes
the sign of $\Omega$, but~$\Pi$ is changed when one puts $B_\phi=
-\Omega^F r/cB_z$ instead of $B_\phi=\Omega^F r/cB_z$. Let us list
all transformations
which do not change the equations for the radial
displacement $\xi_r$ in the perturbations (Istomin~\& Pariev 1996) 
or lead to complex conjugated equations for $\xi_r^\star$:
\begin{enumerate}
\item $\omega\to -\omega^\star$, $\,k\to -k$, $\,\Omega\to -\Omega$;
\item $\omega\to -\omega^\star$, $\,k\to -k$, $\,m\to -m$;
\item $\Omega\to -\Omega$, $\,m\to -m$;
\item $k\to -k$, $\,B_\phi=+\Omega^F r B_z/c\, \to\, B_\phi=-\Omega^F rB_z/c$;
\item $B_z\to -B_z$.
\end{enumerate}
So, if the eigenvalue stability problem for the radial 
displacement has one set of parameters listed above as a solution, 
then the parameters transformed according to the rules 1-5 above 
will also be a solution of the eigenvalue problem for the radial displacement 
for the same~$\xi_r$ or for~$\xi_r^\star$. 
Therefore, $k_{\rm min}$ and $\omega_{\rm min}$ can be transformed according to the 
rules 1-5 above and the transformed $k_{\rm min}$ and $\omega_{\rm min}$ will also correspond 
to the ``standing wave'' wavenumber and frequency. 
The value of~$r_A$ is also not changed under
transformation of the parameters listed above. 
The degree of polarization~$\Pi$ is neither influenced
by changing the sign of~$B_z$ nor by changing sign of~$\Omega$.
The crucial fact is that a sign change in the relation $B_\phi=\pm \Omega^F r B_z/c$
leads to a sign change of $v_{\rm phas}=\omega_{\rm min}/k_{\rm min}$, whilst all other
four transformations listed above change neither~$\Pi$ nor $v_{\rm phas}$.
Therefore, for the purpose of finding the correlation between~$\Pi$ and $v_{\rm vis}$, which
can be verified observationally, all different cases are reduced to considering
two signs in the relation $B_\phi=\pm \Omega^F r B_z/c$, but we can limit
ourself only to the consideration of 
$B_z>0$, $m>0$ and all $\Omega$ for which $r_A$ is located
inside the jet. Using formulae~(\ref{eqn44})--(\ref{eqn45c}) 
and expressions~(\ref{eqn3c}) for $u_z$ and $u_\phi$ one can check that
the Stokes parameters remain unchanged when one changes the sign of $B_\phi$ and
replaces angle $\theta$ by $\pi-\theta$. Also, a simultaneous sign change 
of $v_{\rm phas}$ and change of $\theta$ to $\pi-\theta$ leads to $v_{\rm vis}$
being transformed into $-v_{\rm vis}$. Therefore, the dependence of $\Pi$ on $v_{\rm vis}$
in the case $B_\phi=-\Omega^F r B_z/c$ can be obtained from the same dependence
in the case $B_\phi=+\Omega^F r B_z/c$ by reversing the sign of 
$v_{\rm vis}\to -v_{\rm vis}$
and relabelling each curve from $\theta$ to $\pi-\theta$.

\begin{figure}
\centering
\includegraphics[width=0.3\textwidth]{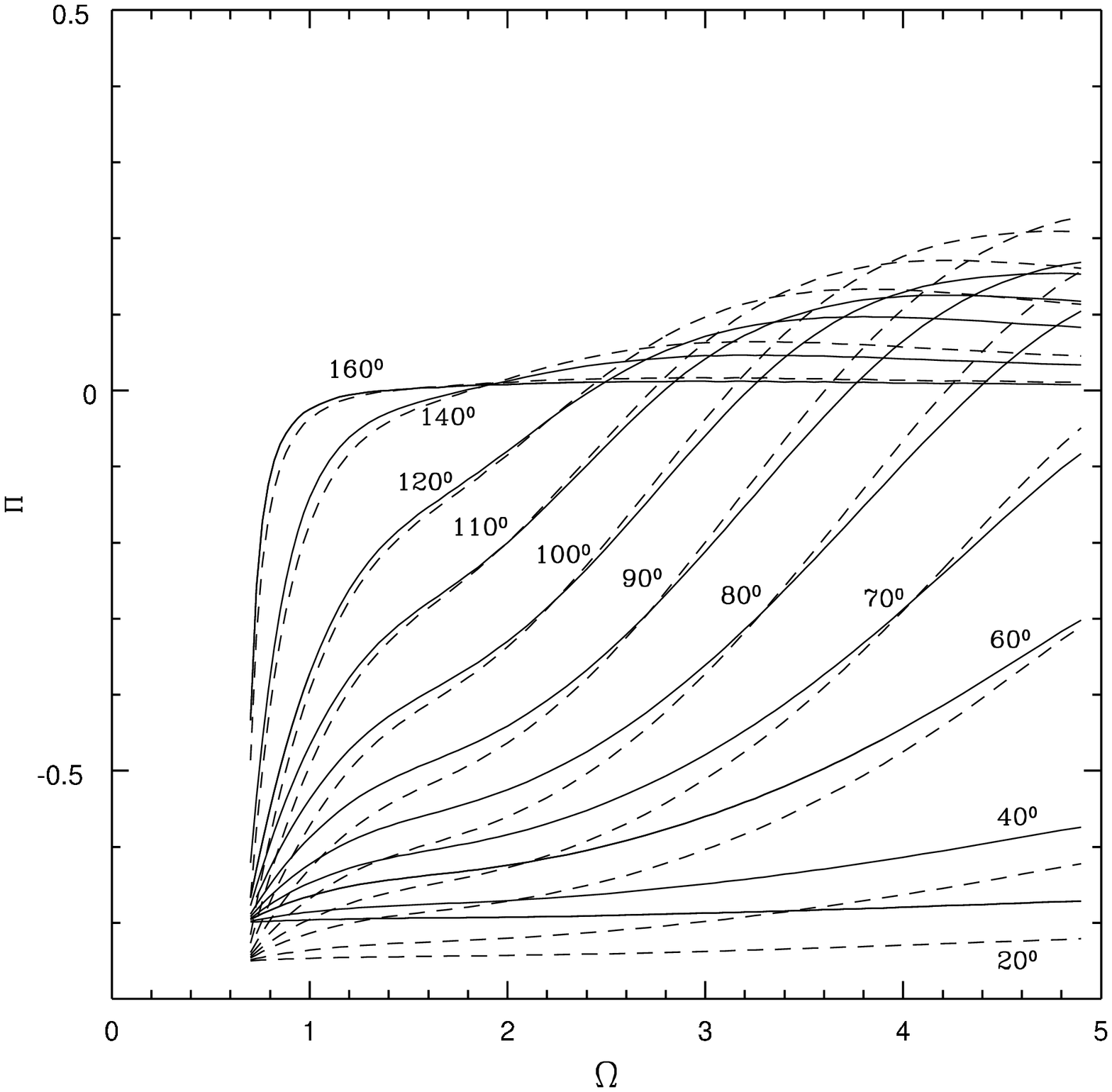}
\hfill
\includegraphics[width=0.3\textwidth]{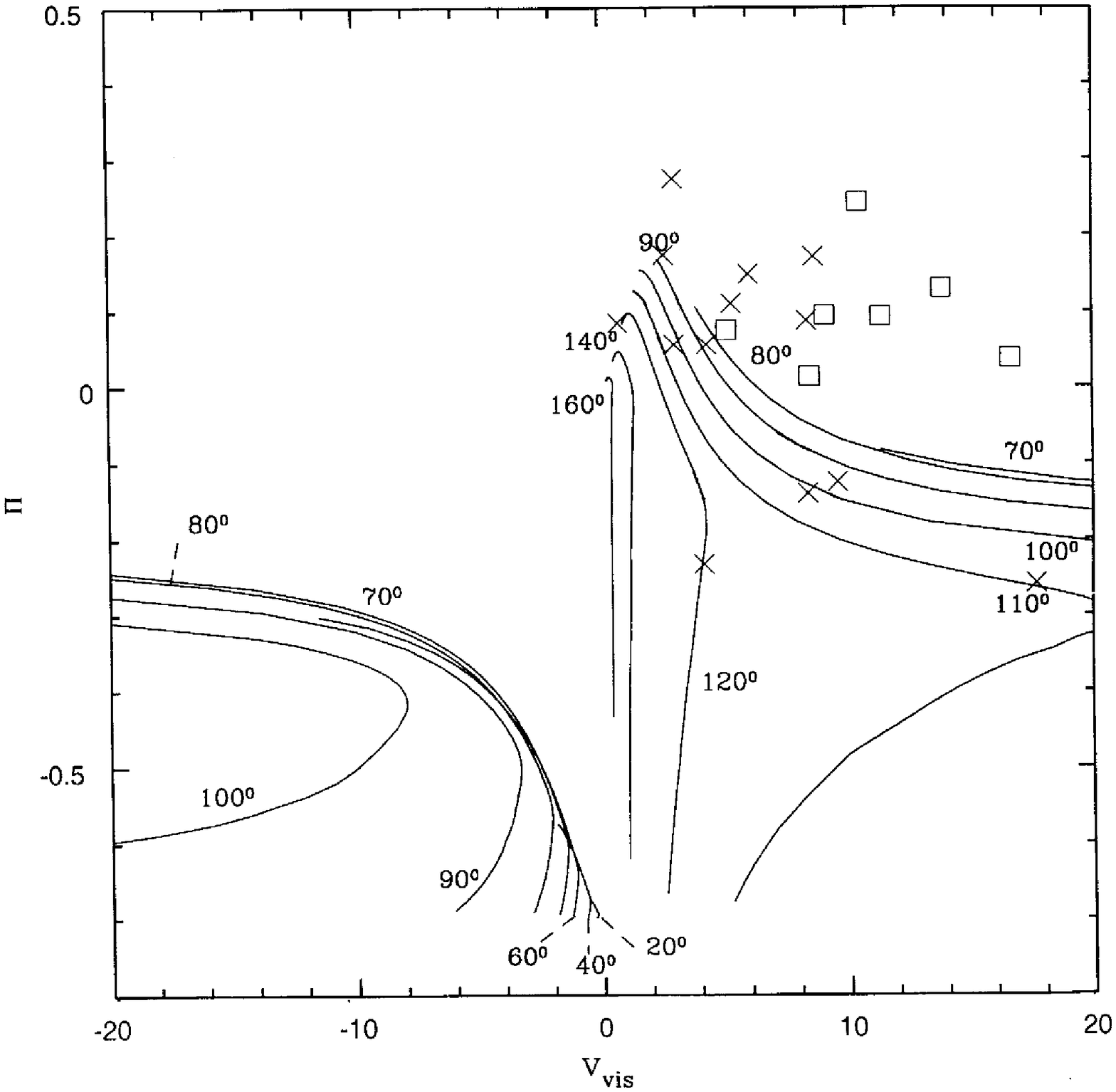}
\hfill
\includegraphics[width=0.3\textwidth]{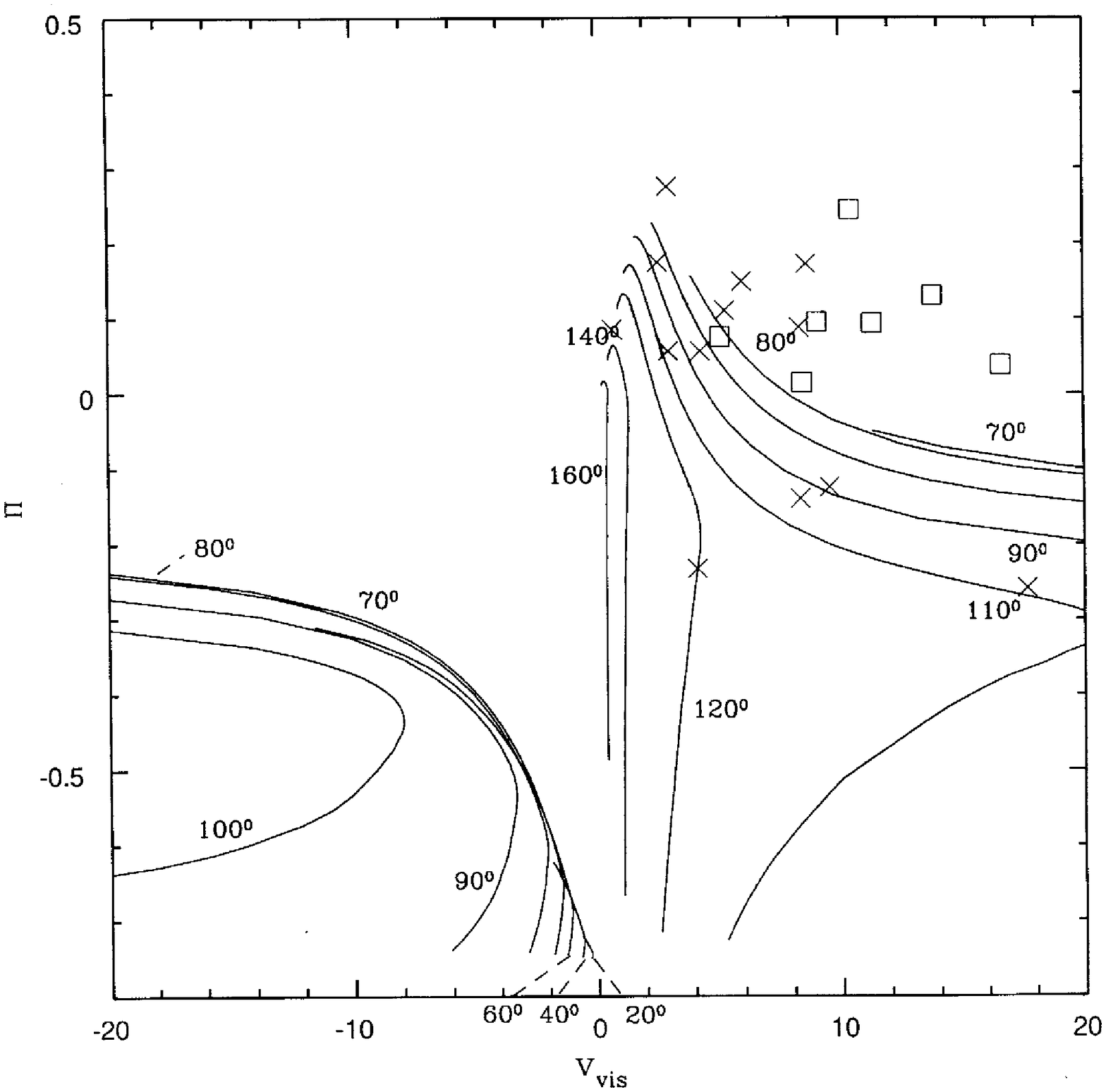}
\caption{The dependence of polarization $\Pi$ on $\Omega$ (left panel), 
on $v_{\rm vis}$
for $\g=2.1$ (middle panel) and for $\g=3$ (right panel). Emitting particles are
concentrated close to the Alfv\'en resonant surface.
$B_\phi=-\Omega^F r B_z/c$, fundamental eigenmode with $m=1$.
Notations are the same as in Fig.~\protect\ref{fig1}. In the left panel
solid curves are for $\g=2.1$, dashed curves are for $\g=3.$}
\label{fig4}
\end{figure}

We plotted the degree of linear polarization~$\Pi$ versus~$v_{\rm vis}$ for $m=1$ in
Figs.~\ref{fig3} and~\ref{fig4} for $B_\phi=+\Omega^F r B_z/c$
and $B_\phi=-\Omega^F r B_z/c$
correspondingly. The Alfv\'en resonance exists for $0.7<\Omega<4.9$. Therefore,
the value of $\Omega$ belongs to this interval 
along the curves in Figs.~\ref{fig3} and~\ref{fig4}. 

We see that the curves in Fig.~\ref{fig3} do not match
the observational data for
knots in jets, while those in Fig.~\ref{fig4} do. The dependence on the power-law
index~$\g$ of emitting particles is weak. 
Our choice of $\g
=2.1$ and $\g=3$ is motivated by the maximum and minimum values of $\g$
obtained in the process of the formation of the spectrum of the particles 
accelerated near the Alfv\'en resonance surface (Beresnyak et al. 2002).

We can suggest the following explanation of the fact that knots are observed
more often in jets with $B_\phi=
-\Omega^F r B_z/c$  than in jets having the opposite
winding of the magnetic field. If we look at the models of the formation
of magnetically driven jets from accretion discs (see, e.g.~Pelletier~\&
Pudritz 1992) we
find that the magnetic field lines are always bent in the direction opposite
to the direction of disc rotation as soon as the velocity of the outflow is
directed outwards from the disc plane. This means that in the jet
approaching the observer $B_\phi=-\Omega^F r B_z/c$, whilst in the jet receding
from the observer $B_\phi=+\Omega^F r B_z/c$. Disturbances propagating along
the approaching jet should be brighter than those in the counter-jet due to
the Doppler boosting effect. Therefore, we will observe knots in parts
of jets having $B_\phi=-\Omega^F r B_z/c$ more often than in those parts where
$B_\phi=+\Omega^F r B_z/c$. This can explain the fact that the number of sources
having $\Pi>0$ is somewhat larger than the number of sources having
$\Pi<0$ despite the fact that for most of the values of the parameters calculations give
a longitudinal polarization $\Pi<0$. We see from Fig.~\ref{fig4} that our model
curves fit the data for BL~Lac objects better than those for quasars. On the other hand,
the polarization measurements for the quasars are quite different at different 
frequencies and such differences are larger than the variations of the polarization
measurements for the BL~Lac objects. In the frame of our model, jets in quasars tend 
to have higher values of~$\Omega^F$ than do jets in BL~Lac objects. 
It is remarkable that in quasars knots in jets also have  faster apparent motions
than in BL~Lac objects.

\section{Summary}\label{sec4}

We calculate the linear polarization of the synchrotron radiation produced by 
fast particles in non-uniformly moving and differentially rotating 
relativistic flow such as jet flows in AGNs.
The magnetic field in the jet
has axial and toroidal components. The distribution of the fast particles is 
taken to be isotropic in the reference frame comoving with the flow.
We perform integration of Stokes parameters 
$I$, $U$, $Q$ along the line of sight 
passing through the jet 
with all relativistic effects fully taken into account.
We consider two possibilities for the spatial density of the fast particles:
1)~ultra-relativistic
particles distributed uniformly across the jet, 2)~ultra-relativistic particles
concentrated close to the Alfv\'en resonance surface. 
The latter distribution
follows from our acceleration theory and we compare it with the former
which might be produced by other acceleration processes. We assume a uniform
axial magnetic field in the jet, so the equilibrium configuration of the jet 
depends only on the radial profile of the angular rotational
velocity $\Omega^F(r)$ of the magnetic field lines. 
We take the 
profile of the angular rotational velocity to be a plausible 
monotonic function of radius,
$\displaystyle\Omega^F=\Omega \frac{c}{R}\left(1-\left(\frac{r}{R}
\right)^2\right)$, and investigate the dependence of polarization 
on the magnitude of the angular velocity $\Omega$. 
In our computations we also specify
a component of the velocity parallel to the magnetic field lines such that
the kinetic energy associated with the flow is minimized. 
The actual velocity is determined by the process
of jet formation, which is poorly understood now; still such minimal velocity
is physically favoured~(Frolov~\& Novikov 1998). We
calculate the degree of linear polarization and orientation of the electric 
vector with respect to the jet axis for the emission integrated over the jet cross
section, taking into account the relativistic bulk spiral flow of plasma
in the jet. This flow leads to the presence of an electric field comparable in
magnitude with the magnetic field, causes a Doppler boost of the intensity
of the synchrotron radiation, and a swing of the polarization position angle.
We want to stress that the last effect, which was first pointed out in 1979 by
Blandford~\& K\"onigl for the
simpler case of a uniform magnetic field parallel to the jet axis, 
substantially modifies the resulting degree of
linear polarization from the whole jet. However, because of the axial symmetry
of the problem, only two orientations of polarization position angles are
possible: parallel or perpendicular with respect to the jet axis.
Observations of polarization of VLBI scale jets in BL~Lac objects 
on centimetre wavelengths (Gabuzda et al., 2000) show a bimodal distribution 
of the position angles: most of the knots in
BL~Lac objects have polarization either parallel to the jet axis
or perpendicular to it. Similar observations of quasars (Lister 2001)
on millimetre wavelengths
indicate a unimodal distribution of the position angles: most of the knots 
in the observed quasars have the electric vector of polarized emission oriented 
along the jet axis. In our present calculations the polarization does not depend on 
frequency since we have assumed a power-law distribution over the whole range of energies 
of the particles. Different slopes of the energy spectrum of the particles
at different energies can result in a dependence of the degree of polarization 
and of the position angle on the frequency. 
This may account for the differences in polarizations at  different 
frequencies observed in quasars. At present, we can only suggest that jets 
in quasars tend to have larger $\Omega^F$ and a larger ratio of toroidal to
poloidal magnetic field than do jets in BL~Lac objects. This can account for 
the electric field vector in quasars being predominantly parallel
to the jet axis.

We also calculated the velocities of crests of standing eigenmodes and
compared them with the observed proper motions of bright VLBI knots in jets.
We were able to match the observational data only for the case when the emitting
particles are concentrated close to the Alfv\'en resonance surface and the
magnetic field lines are twisted in the direction opposite to the
direction of rotation $\Omega^F$. The last fact can be understood by a selection
effect due to the Doppler boosting which makes jets approaching
the observer more favourable to observe than receding jets.
Comparing  calculations with observations we can estimate the angular
rotational velocity of the jets. The trend is that $\Omega^F$ in BL~Lac objects
is intrinsically smaller than in quasars. 
More detailed calculations considering 
different possibilities for the $\Omega^F(r)$ profile are postponed 
to another work. There are some discrepancies between our calculations   
and the observation points (see Fig.~\ref{fig4}). It is possible that the radial 
profile of $\Omega^F(r)$ needs to be tuned to provide better agreement.
At present it is difficult to suggest a unique test, which would 
make it possible to distinguish between electromagnetic and shock mechanisms  
for the origin of knots in jets.  
Future high-resolution and 
high-dynamic-range observations could provide detailed images 
of the spatial structure of knots and surrounding jets, and provide the answer.

\begin{acknowledgements}
The authors are grateful to D.C. Gabuzda for compiling observational data
on polarization of jets and helpful discussions of different
observational issues.
This work was done under the partial support of the Russian Foundation
for Fundamental Research (grant number 02-02-16762).
Support from DOE grant DE-FG02-00ER54600 is acknowledged.
\end{acknowledgements}

\newpage

\begin{center}
{\Large References}
\end{center}
\parindent0pt

Attridge~J.M., Roberts~D.H., \& Wardle~J.F.C., 1999, ApJ, 518, L87

Begelman~M.C., Blandford~R.D., \& Rees~M.J., 1984, Rev. Mod. Phys., 56, 255 

Beresnyak~A.R., Istomin~Ya.N., \& Pariev~V.I., 2003, A\&A, accepted

Blackman~E.G., 1996, ApJ, 456, L87

Blandford~R.D., \& K\"onigl~A., 1979, ApJ, 232, 34

Blandford~R.D., \& Eichler~D., 1987, Physics Reports, 154, 1

Cocke~W.J., \& Holm~D.A., 1972, Nature Phys. Sci., 240, 161

Frolov~V.P., \& Novikov~I.D., 1998, Black hole physics: basic concepts and 
new developments. Kluwer Academic, Dordrecht 

Gabuzda~D.C., \& Cawthorne~T.V., 1993, in Davis~R.J., Booth~R.S., eds,
Subarcsecond Radio Astronomy. Cambridge Univ. Press, Cambridge, p.211

Gabuzda~D.C., Mullan~C.M., Cawthorne~T.V., Wardle~J.F.C.,
\& Roberts~D.H., 1994, ApJ, 435, 140 

Gabuzda~D.C., 1999a, New Astronomy Reviews, 43, 691

Gabuzda~D.C., 1999b, in Ostrowski~M., Schlickeiser~R., eds,
Plasma Turbulence and Energetic Particles in Astrophysics. 
Uniwersytet Jagiellonski, Cracow, 1999, p.301

Gabuzda~D.C., 2000, in Conway~J.E., Polatidis~A.G., Booth~R.S., Pihlstr\"om~Y.M.,
eds, EVN Symposium 2000, Proceedings of the 5th European 
VLBI Network Symposium. Onsala Space Observatory, Onsala, p.53

Gabuzda~D.C., Pushkarev~A.B., \& Cawthorne~T.V., 2000, MNRAS, 319, 1109

Gabuzda~D.C., Pushkarev~A.B., \& Garnich~N.N., 2001, MNRAS, 327, 1

Ginzburg~V.L., 1989, Applications of Electrodynamics in Theoretical Physics and
Astrophysics. Gordon and Breach Science Publishers, New York, p.65 

Hirotani~K., Iguchi~S., Kimura~M., \& Wajima~K., 1999, PASJ, 51, 263

Hirotani~K., Iguchi~S., Kimura~M., \& Wajima~K., 2000, ApJ, 545, 100

Hutchison~J.M., Cawthorne~T.V., \& Gabuzda~D.C., 2001, MNRAS, 321, 525

Istomin~Ya.N., \& Pariev~V.I., 1994, MNRAS, 267, 629 

Istomin~Ya.N., \& Pariev~V.I., 1996, MNRAS, 281, 1 

K\"onigl~A., \& Choudhuri~A.R., 1985, ApJ, 289, 188

Laing~K.A., 1981, ApJ, 248, 87

Lepp\"anen~K.J., Zensus~J.A., \& Diamond~P.J., 1995, AJ, 110, 2479

Lister~M.L., Marscher~A.P., \& Gear~W.K., 1998, ApJ, 504, 702

Lister~M.L., 2001, ApJ, 562, 208

Lister~M.L., Tingay~S.J., \& Preston~R.A., 2001, ApJ, 554, 964

Manolakou~K., Anastasiadis~A., \& Vlahos~L., 1999, A\&A, 345, 653

Morrison~P., \& Sadun~A., 1992, MNRAS, 254, 488

Pelletier~G., \& Pudritz~R., 1992, ApJ, 394, 117 

Reynolds~C.S., Fabian~A.C., Celotti~A., \& Rees~M.J., 1996, MNRAS,
283, 873.

Romanova~M.M., \& Lovelace~R.V.E., 1992, A\&A, 262, 26

R\"oser~H.-J., \& Meisenheimer~K., 1991, A\&A, 252, 458

Sazonov,~V.N., 1969, AZh, 46, 502

Sikora~M., \& Madejski~G., 2000, ApJ, 534, 109

Scheuer~P.A.G., 1984, Adv. Space Res., 4, 337 

Wardle~J.F.C., Homan~D.C., Ojha~R., \& Roberts~D.H., 1998, Nature,
395, 457

Zhelezniakov~V.V., 1996, Radiation in astrophysical plasmas. 
Kluwer, Dordrecht

\end{document}